\documentclass[12pt,preprint,psfig]{aastex}
\usepackage{psfig}
\newcommand{\etal}{{et al.} }

\newcommand{\xmm}{{\it XMM-Newton} }

\newcommand{\chandra}{{\it Chandra} }

\newcommand{\suzaku}{{\it Suzaku} }
\newcommand{\suzakup}{{\it Suzaku}}

\newcommand{\hetg}{{\it HETGS} }
\newcommand{\hetgp}{{\it HETGS}}

\newcommand{\fekalfa}{{Fe~K$\alpha$} }

\newcommand{\fexxv}{Fe~{\sc xxv} }
\newcommand{\fexxvp}{Fe~{\sc xxv}}

\newcommand{\feklya}{{Fe~{\sc xxvi }~Ly$\alpha$} }

\newcommand{\feklyap}{{Fe~{\sc xxvi}~Ly$\alpha$}}

\newcommand{\lxledd}{$L_{x}/L_{\rm EDD}$ }

\newcommand{\lflux}{$I_{\rm Fe K\alpha}$}

\newcommand{\lineenp}{$E_{\rm Fe K\alpha}$}

\newcommand{\hbeta}{${\rm H}_{\beta}$ }
\newcommand{\hbetap}{${\rm H}_{\beta}$}

\newcommand{\dovname}{{Dov\v{c}iak} }

\newcommand{\fekecen}{$E_{0}$ }
\newcommand{\fekecenp}{$E_{0}$}
\newcommand{\fekflux}{{$I_{\rm Fe~K}$} }

\newcommand{\feksigma}{{$\sigma_{\rm Fe~K}$} }
\newcommand{\feksigmap}{{$\sigma_{\rm Fe~K}$}}
\newcommand{\figindspec}{{Fig.~1} }
\newcommand{\figindspecp}{{Fig.~1}}
\newcommand{\figspectot}{{Fig.~2} }
\newcommand{\figspectotp}{{Fig.~2}}

\newcommand{\figifevsfwhm}{{Fig.~6} }
\newcommand{\figifevsfwhmp}{{Fig.~6}}
\newcommand{\figecenhist}{{Fig.~3} }
\newcommand{\figecenhistp}{{Fig.~3}}

\newcommand{\figewhist}{{Fig.~4} }

\newcommand{\figewhista}{{Fig.~4(a)} }

\newcommand{\figewhistb}{{Fig.~4(b)} }

\newcommand{\figewhistc}{{Fig.~4(c)} }
\newcommand{\figewhistcp}{{Fig.~4(c)}}
\newcommand{\figewhistd}{{Fig.~4(d)} }
\newcommand{\figewhistdp}{{Fig.~4(d)}}
\newcommand{\figfwhmvshbeta}{{Fig.~5} }
\newcommand{\figfwhmvshbetap}{{Fig.~5}}
\newcommand{\figbaldwin}{{Fig.~7} }

\newcommand{\figbaldwina}{{Fig.~7(a)} }

\newcommand{\figbaldwinb}{{Fig.~7(b)} }

\newcommand{\figbaldwincp}{{Fig.~7(c)}}

\newcommand{\figbaldwindp}{{Fig.~7(d)}}

\newcommand{\tablesigfixed}{Table~1 }
\newcommand{\tablesigfixedp}{Table~1}
\newcommand{\tablesigfree}{Table~2 }
\newcommand{\tablesigfreep}{Table~2}
\newcommand{\tablelinemeans}{Table~3 }
\newcommand{\tablelinemeansp}{Table~3}
\newcommand{\tablebaldwin}{Table~4 }
\newcommand{\tablebaldwinp}{Table~4}

\newcommand{\blue}[1]{{\bf }}

\shorttitle{CHANDRA HEG OBSERVATIONS Fe K$\alpha$ LINES IN AGN}

\begin{document}

\title{THE CORES OF THE Fe K$\alpha$ LINES IN ACTIVE GALACTIC NUCLEI: AN
EXTENDED Chandra HIGH ENERGY GRATING SAMPLE}

\author{X. W. Shu\altaffilmark{1,2}, T. Yaqoob\altaffilmark{2},  J. X. Wang\altaffilmark{1}}

\altaffiltext{1}{
CAS Key Laboratory for Research in Galaxies and Cosmology, 
Department of Astronomy, University of Science and Technology of China, 
Hefei, Anhui 230026, P. R. China, xwshu@mail.ustc.edu.cn}
\altaffiltext{2}{Department of Physics and Astronomy,
Johns Hopkins University, Baltimore, MD 21218, yaqoob@skysrv.pha.jhu.edu}

\begin{abstract}

We extend the
study of the core of the \fekalfa emission line at $\sim 6.4$~keV
in Seyfert galaxies reported in Yaqoob \& Padmanabhan (2004)
using a larger sample observed by the \chandra High Energy Grating (HEG). 
The sample consists of 82 observations of 36 unique sources
with $z<0.3$.
Whilst heavily obscured active galactic nuclei (AGNs)
are excluded from the sample,
these data offer some of the highest precision measurements
of the peak energy of the \fekalfa line, and the
highest spectral resolution measurements of the
width of the core of the line in unobscured and moderately
obscured ($N_{H}<10^{23} \ \rm cm^{-2}$) Seyfert galaxies
to date. From an empirical and uniform analysis, we
present measurements of the \fekalfa line centroid energy,
flux, equivalent width (EW), and intrinsic width (FWHM).
The \fekalfa line
is detected in 33 sources, and its centroid energy
is constrained in 32 sources. 
In 27 sources the statistical quality of the data 
is good enough to yield measurements of the FWHM. We
find that the distribution in the line centroid energy
is strongly peaked around the value for neutral Fe,
with over 80\% of the observations giving values in the
range 6.38--6.43~keV. Including statistical errors,
30 out of 32 sources ($\sim 94\%$) have a 
line centroid energy in the range 6.35--6.47~keV. The mean equivalent width,
amongst the observations in which a non-zero lower limit
could be measured, was $53 \pm 3$~eV.
The mean FWHM from the subsample of 27 sources was 
$2060 \pm 230 \ \rm km \ s^{-1}$. { The mean EW and FWHM are
somewhat higher when multiple observations for a given source
are averaged.}
From a comparison
with the H$\beta$ optical emission-line widths (or, for one source,
Br$\alpha$), we find that there is no universal location
of the \fekalfa line-emitting region relative to the optical BLR.
In general, a given source may have contributions to the \fekalfa line
flux from parsec-scale distances from the putative black hole,
down to matter a factor $\sim 2$ closer to the black hole than the BLR.
We confirm the presence of the X-ray Baldwin effect,
an anti-correlation between
the \fekalfa line EW and X-ray continuum luminosity. The HEG data
have enabled isolation of this effect to the narrow core of the \fekalfa line.
\end{abstract}


\section{INTRODUCTION}
\label{intro}

The narrow core 
(FWHM~$<10,000 \ \rm km \ s^{-1}$) 
of the \fekalfa fluorescent emission, peaking
at $\sim 6.4$~keV is a common and dominant feature of the X-ray
spectrum of active galactic nuclei (AGNs) that
have a 2--10~keV X-ray luminosity less than $\sim 10^{45} \rm \ erg \ s^{-1}$
(e.g. Sulentic \etal 1998; Lubi\'{n}ski \& Zdziarski 2001;
Weaver, Gelbord, \& Yaqoob 2001; 
Perola \etal 2002; Yaqoob \& Padmanabhan 2004 (hereafter YP04); 
Levenson \etal 2006;
Winter \etal 2009). The luminosity in the core of the
\fekalfa emission may be comparable to any additional,
relativistically-broadened \fekalfa line emission that may be
present, and indeed, in many cases may be the {\it only}
component of the \fekalfa line (e.g. see 
Guainazzi, Bianchi, \& \dovname 2006;
Nandra \etal 2007; Miller 2007; Turner \& Miller 2009; Bianchi \etal 2009).
Measurement of the properties of the core of the \fekalfa
line in AGN is important for two principal reasons. One is
to constrain the physical properties of the large-scale structure
in the central engine. 
The peak energy of the \fekalfa line constrains the
ionization state of the line-emitting matter, and the width
of the line gives kinematic information that can be
used to estimate the size and location of the X-ray reprocessor.
The equivalent width (EW) of the \fekalfa line is a function
of geometry, column density, covering factor,
element abundances, and orientation of the 
line-emitter.
Another reason why spectroscopy of the \fekalfa line
core is important is that it is necessary
to model the narrow component of the line in order to
deconvolve any relativistically-broadened 
emission-line component that may be present.
The \chandra high energy grating (HEG;
see Markert \etal 1995) is still unsurpassed in spectral
resolution in the Fe~K band, which at 6.4~keV is
$\sim 39$~eV, or $\sim 1860 \ \rm km \ s^{-1}$ FWHM.
This is a factor of $\sim 4$ better than the spectral
resolution of X-ray CCD detectors aboard \xmm and \suzakup.
Although broad \fekalfa emission lines are better studied
with CCD spectrometers (due to their higher throughput), the
\chandra HEG is well-suited for studying the narrow core
of the \fekalfa line. One can then utilize the HEG measurements
to deconvolve narrow and broad \fekalfa line components
in lower spectral resolution data.

In YP04 the results of
a uniform analysis of the properties of the \fekalfa
emission-line core were presented, based on \chandra
HEG data of a modest sample
of fifteen AGN. There are now a larger number of
\chandra HEG observations for which the data are available,
and in the present paper we extend the study of YP04
to include 82 observations of 36 unique AGN.
The paper is organized as follows. In \S\ref{data} we
describe the observations and data. In \S\ref{hegspec}
we describe the methodology and basic spectral-fitting
results. In \S\ref{properties} we discuss the implications
of the results for the properties of the core of narrow
\fekalfa emission line in the HEG AGN sample. In \S\ref{baldwineffect}
we investigate whether the narrow core of the
\fekalfa line as isolated by the HEG, supports the so-called
X-ray Baldwin effect (an anti-correlation 
between the line EW and X-ray luminosity and between the  
EW and a proxy for the accretion rate).
In \S\ref{summary} we summarize our results and findings.

\section{OBSERVATIONS AND DATA}
\label{data}
 
The \chandra 
\hetg ({\it High Energy Transmission Grating Spectrometer})
consists of two grating assemblies,
a High-Energy Grating (HEG) and a Medium-Energy Grating (MEG),
and it is the HEG that achieves the highest spectral resolution.
The MEG has only half of the spectral resolution
of the HEG and less effective area in the Fe~K band, so our study will
focus on the HEG data. 
Our study is based on data from \chandra \hetg AGN observations
that were public as of 2008, September 30, filtering on several
criteria. Firstly, we selected non-blazar AGN that had
$z<0.3$. This actually only omitted one source, PKS~2149$-$306
($z=2.345$), which is a high-luminosity radio-loud quasar 
(see Fang \etal 2001 for results from the \chandra grating observations).
{ Since
the centroid energy of the \fekalfa line appears at $\sim 6.4/(1+z)$~keV,
the line would appear at very different places on the instrumental
effective area curve for very different values of $z$. In addition,
the EW of the \fekalfa line is smaller by a factor $(1+z)$ compared
to the rest-frame value. Therefore, a restriction on the sample redshift
also helps to achieve a more homogeneous analysis. 
}Next we 
required that the total counts 
in the full HEG bandpass ($\sim 0.9-8$~keV) was $>1500$,
a condition which rejects spectra that have insufficient
signal-to-noise ratios for our purpose. Relaxing this
criterion would only have admitted two sources, PG~1404$+$226,
and 1H~0707$-$495.
We then selected those AGN that are known to have
X-ray absorbing column densities less than $10^{23} \ \rm 
\ cm^{-2}$. The reason for this is that AGN with higher
column densities have X-ray spectra that are complex and
measurements of the properties of even the narrow \fekalfa line core
in such sources can become model-dependent. Indeed,
Murphy \& Yaqoob (2009; hereafter MY09) showed, using monte carlo simulations
of X-ray reprocessing, that
inclination-angle and geometrical effects on the EW of the 
\fekalfa line become important for column densities
greater than $\sim 10^{23} \ \rm  \ cm^{-2}$. 
Although the column density out of the line-of-sight could
be larger than the line-of-sight column density, it is the
simplicity of the observed spectrum that is the driver
of the selection. 
We will present a study of
heavily-absorbed AGN observed by the \chandra HEG elsewhere.
Our approach in the present paper is to perform
a very simple {\it empirical} analysis in order to obtain robust
measurements of the basic narrow \fekalfa line core
parameters that are not dependent
on details of how the continuum is modeled. 
Our selection criteria then populate our sample
with some sources that are
formally classified as type~2 AGN, whereas the study of YP04
included strictly only type~1 AGN.

We also excluded 15 $Chandra$~\hetg observations of M81 ($\sim$435 ks exposure) 
from the study, as its 
very low luminosity and accretion rate
set it apart from the rest of the bright AGN in the sample
(it is most often classified as a LINER harboring a low-luminosity AGN).
The bolometric luminosity of M81 is only $\sim10^{-5}$ of the
Eddington luminosity (Young \etal 2007).
We note that the results of some \hetg results for M81,
based on $\sim 280$~ks exposure time have been presented
by Young \etal (2007) who found K-shell emission lines from He-like and
H-like Fe in addition to the \fekalfa line at $\sim 6.4$~keV.
Our final sample consists of 82 observations of 36 unique AGN
and includes all of the observations in YP04
(which we re-analyzed for the present paper).
We note that our sample includes 3C~273, which is sometimes
classified as a blazar. However, this source is variable
and often shows Seyfert-like properties (e.g. Grandi \& Palumbo 2007).

The \chandra data for the sample 
were reduced and HEG spectra were made 
as described in Yaqoob \etal (2003) and YP04.
We used only the first orders of the grating data (combining
the positive and negative arms). 
The mean HEG count rates ranged from $0.026 \pm 0.001$ ct/s
for the weakest source (PDS 456) to $1.161 \pm 0.006$ for
the brightest source (IC 4329a).
The exposure time ranged from $\sim $20 ks to $\sim $172 ks
per observation, but
was $\sim 50-120$ ks for most of the sources.
Nineteen sources were observed more than once, and 
the largest net exposure time from summed data from
observations of the
same source was $\sim 880$~ks (NGC~3783).
The observations, identified by target name, sequence number,
and observation ID (``ObsID''), are listed in \tablesigfixedp, along with
the net exposure times for the spectra.
Further details of all of the observations can be found
in the \chandra data archive
at {\tt http://cda.harvard.edu/chaser/}. Higher-level products,
including lightcurves and spectra for each observation
can be found in the databases {\it HotGAS} 
({\tt http://hotgas.pha.jhu.edu}), and {\it TGCat}
({\tt http://tgcat.mit.edu/}).
Background was not subtracted since it is negligible over the
energy range of interest (e.g. see Yaqoob \etal 2003).
Note that 
the systematic uncertainty in the HEG
wavelength scale 
is $\sim 433 \ \rm km \ s^{-1}$ ($\sim 11$~eV) at 6.4 keV
\footnote{http://space.mit.edu/CXC/calib/hetgcal.html}. 

\section{SPECTRAL FITTING METHODOLOGY AND RESULTS}
\label{hegspec}

The spectra were analyzed
using the spectral-fitting package XSPEC (Arnaud 1996).
Since we are interested in utilizing the highest possible
spectral resolution available, we used spectra binned
at $0.0025\AA$, and this amply oversamples the HEG resolution
($0.012\AA$ FWHM).
The $C$-statistic was used for minimization.
All model parameters will be
referred to the source frame.
Our method is simply to fit a simple continuum plus Gaussian
emission-line model over the 2--7~keV band for each spectrum.
Above 7~keV the HEG effective area rapidly decreases.
{ We found, as in YP04, that if the energy band is
restricted any further the constraints on the \fekalfa
line parameters do not improve because when the
intrinsic line width is free there is degeneracy of the
line parameters with the continuum slope.}
In most cases a simple power-law continuum was adequate, but
for some sources an additional uniform, neutral, absorber
component was included (namely NGC 2110, 
MCG -5-23-16, NGC 4151 and NGC 5506).
In no case was a column density 
greater than 4.3$^{+0.4}_{-0.3}\times10^{22}$ cm$^{-2}$ required.
Galactic absorption was not included for any of the sources because 
such small column densities have little effect above 2~keV.
Thus, there were a maximum of { six free parameters in the
model, namely the power-law slope and its overall normalization}, 
$\Gamma$, the column
density, $N_{H}$, the centroid energy of the Gaussian 
emission-line component, $E_{0}$, its flux, $I_{\rm Fe~K}$, and
its width, $\sigma_{\rm Fe~K}$.
The approach of using an over-simplified
continuum model is necessitated by the limited bandpass
of the HEG data ($\sim 2-7$~keV) but since we are
interested in the narrow core of the \fekalfa emission
line, at the spectral resolution of the HEG, this is
not restrictive. Obviously, use of such an empirical model
means that we should not assign a physical meaning to
$\Gamma$ and $N_{H}$.

The signal-to-noise ratio of the
spectra showed a wide range so we followed a systematic, two-step
procedure that accounts for this. In the first round
of analysis we fixed the emission-line width, $\sigma_{\rm Fe~K}$, at 
1 eV (corresponding to
$\sim 110 \ \rm km \ s^{-1}$ FWHM at 6.4~keV), 
a value well below the instrument resolution, because
the line width could not be constrained in all the data sets.
Uniformly freezing the line width for all the data sets then
picks up the narrowest, unresolved core component of the emission line for
all the data sets. 
In the second round of analysis we allowed the line width to float.
Where multiple observations of a given source were
available we constructed
and fitted spectra that were averaged over all of the observations,
in addition to fitting data 
from the individual observations. Inter-observation
variability will be discussed in \S\ref{lineflux}. 

The results of this first round of analysis are shown in
\tablesigfixed which shows the derived equivalent width (EW)
in addition to the other \fekalfa line
parameters. 
Note that since the models were fitted
by first folding through the instrument response before comparing
with the data, the derived line parameters {\it do not}
need to be corrected for instrumental response.
We do not give the best-fitting values of $\Gamma$ or $N_{H}$
in \tablesigfixed because the values derived using the simplistic
continuum model are not physically meaningful but are
simply parameterizations.
The 2--10~keV continuum fluxes and luminosities shown in
\tablesigfixed were obtained by extrapolating the best-fitting model
up to 10~keV. {We caution that such extrapolation could
give inaccurate fluxes and luminosities if the continuum shape
is significantly different in the 7--10~keV band compared
to the extrapolated model.}
The fluxes are not corrected for absorption,
but the luminosities are. 
The $\Delta C$ values
shown in \tablesigfixed correspond to the 
decrease in $C$ when the narrow, two-parameter
emission line was added to the continuum-only model,
and is therefore a measure of the significance of the emission line.

We found that in some cases
(fourteen observations of ten sources,
plus the summed spectrum of IRAS~$18325-5926$) the \fekalfa line
centroid energy could not be constrained, and in such cases
the centroid energy was fixed at 6.40~keV. In twelve of these
data sets the \fekalfa line was not detected at a confidence
level greater than 90\% and for these cases only an upper
limit on the EW, and line flux, $I_{\rm Fe~K}$, could be obtained.
Two sets of statistical errors are given for the \fekalfa line
parameters in \tablesigfixed for each spectral fit. The
first set corresponds to 68\% confidence 
($\Delta C =  2.279$, or 0.989, depending on whether 
there were two parameters or one parameter free respectively in 
the Gaussian component). These ``1-$\sigma$'' errors are
useful for performing standard statistical analyses on the
model parameters. However, as a more conservative measure,
the 90\% confidence {\it range} (for the appropriate number
of free parameters of the Gaussian component)
for each line parameter is also given 
in \tablesigfixed (values in parentheses). For the 90\% confidence
ranges $\Delta C =  4.605$ and 2.706 for two parameters and one
parameter free respectively.

We also found that in
some sources that have multiple observations,
the \fekalfa line parameters were sometimes better constrained
from some of the individual observations than from the
averaged spectra because the latter may contain contributions
from data in which the \fekalfa line was relatively weaker
in EW. 
Detailed interpretation of the results in \tablesigfixed will be given in
\S\ref{properties}.

In the second round of spectral fitting we allowed the 
intrinsic width of the Gaussian emission-line component to
be a free parameter. 
However, in situations when the signal-to-noise ratio
of the \fekalfa emission line is too poor, the Gaussian 
model emission-line component
can become very broad as
it then begins to model the continuum, resulting in values of the
width that are not actually related to the physical width of the
emission line. 
As a very loose initial criterion, we rejected
all cases in which a three-parameter Gaussian component
was detected with less than 95\% confidence (corresponding
to rejecting fits that gave $\Delta C<7.8$). This rough
criterion immediately rejected fits for which the 
fits actually became unstable and
left 26 unique sources and 65 data sets, 
including 14 spectra averaged over multiple observations.
The results
for all of the fits with $\Delta C>7.8$ are shown in \tablesigfreep,
in which the three-parameter, 68\% statistical errors and 90\%
confidence ranges on the line parameters are given.
The \fekalfa line width is given as
a FWHM in $\rm km \ s^{-1}$ rather than the Gaussian
width, $\sigma_{\rm Fe~K}$.

The next selection criterion
we used was more specifically focussed on determining whether the
model width was a measure of the true line width.
Owing to the excellent spectral resolution of the HEG
it is straightforward to determine when the model FWHM is no
longer a measure of the \fekalfa emission-line width by reconciling
the spectral data with the confidence contours of
\lflux versus the centroid energy, \lineenp.
This can be seen in \figindspec and \figspectot which show,
for a given source,
the rest-frame spectra in the Fe~K region alongside the
confidence contours of \lflux versus \lineenp.
We have not shown plots for all the data sets (spectra for all of the
data sets can be found in the databases mentioned in \S\ref{intro}).
For example, we have not shown plots for
data sets that have already been presented in YP04.
Nor have we shown plots for data sets in which the detection of
the \fekalfa line was marginal or insignificant. 
\figindspec shows results for
sources that have only one observation whilst
\figspectot shows time-averaged spectra for sources with multiple
observations, alongside confidence contours for the individual
and time-averaged data. 
We see in \figindspec and \figspectot that in some cases the
99\% confidence contours indicate a range in centroid energy
that is clearly much larger than the breadth of the emission-line
feature that can be estimated directly from the spectral plots.
For example, for NGC~985 the
joint, two-parameter, 
99\% confidence  contour of line intensity versus
center energy (solid line) is $\sim$600 eV wide.
However, from the spectral plot
in the Fe K band, the \fekalfa line clearly has a width less than $\sim250$ eV.
Thus the 99\% confidence bounds on line flux versus
line centroid energy imply that the 
Gaussian component is in fact modeling the
continuum, resulting in very large
FWHM values that are not actually related to the physical width
of the emission line.
For this case of NGC~985, 
we constructed a 99\% confidence contour (dotted line in \figindspecp)
of the line intensity versus energy
with the Gaussian width fixed at 1~eV 
($ 110 \rm km \ s^{-1}$). This
shows that the centroid energy is constrained 
to be in the range $\sim$ 6.3 -- 6.5 keV,
consistent with the physical width of the narrow core in the spectral plot.

Thus, by comparing the line intensity versus
line energy confidence contours with the spectra
we determined that the FWHM constraints deduced from
spectral fits for 16 out of the 65 data sets
were not reliable indicators of the \fekalfa line
core intrinsic width (none of the 16 are data sets summed over
multiple observations). 
We found that the situations in which the Gaussian width
model parameter became an unreliable indicator of the 
emission-line intrinsic width generally corresponded to
a 90\% confidence, two-parameter upper limit on the FWHM greater 
than $\sim 15,000 \rm \ km \ s^{-1}$. 
We note that even for the cases where we can obtain
a reliable measure of the \fekalfa line FWHM, the true
line width may be less than the FWHM deduced from our 
simplistic model-fitting because there may be 
blending from an unresolved Compton-shoulder component and/or
from several (low) ionization states of Fe. Also, the 
difference in rest energy of $\sim 13$~eV of the individual
components of the
\fekalfa line ($K\alpha_{1}$ and $K\alpha_{2}$)
may increase the apparent FWHM if the
line is modeled with a single Gaussian model component. However, the
separation of $K\alpha_{1}$ and $K\alpha_{2}$
corresponds to $\sim 600 \rm \ km \ s^{-1}$
(three times less than the HEG resolution in the Fe~K band),
and considering the $K\alpha_{1}$:$K\alpha_{2}$ branching
ratio of 2:1, Yaqoob \etal (2001)
showed that artificial broadening is not a concern
for line parameters and signal-to-noise ratio that are typical of
the HEG observations reported here.

In the present paper we are concerned only with the
\fekalfa line core centered at $\sim 6.4$~keV, and not
emission lines from highly ionized species of Fe.
Nevertheless, overlaid on the spectra in \figindspec are
vertical dashed lines marking the positions of
the \fexxv He-like triplet lines (the two intercombination lines
are shown separately), \feklyap, Fe~{\sc i}~$K\beta$, and the
Fe~K-shell threshold absorption-edge energy. The values
adopted for these energies were from 
NIST\footnote{http://physics.nist.gov/PhysRefData} (He-like triplet);
Pike \etal 1996 (\feklyap); Palmeri \etal 2003 (Fe~{\sc i}~$K\beta$),
and Verner \etal 1996 (Fe~K edge).
Emission lines {\it and}
absorption lines from highly ionized species of Fe have indeed
been reported in the literature for some of the 
same data sets discussed in the present paper (e.g. see Bianchi \etal
2005). We summarize such results from the literature
in the appendix for each source as appropriate, including any previous results
on the 6.4~keV emission line that are based on the same data that
we have employed. We also give in the appendix any unusual
details and/or issues for particular data sets  
that are pertinent to our analysis of the \hetg data. 

\section{PROPERTIES OF THE CORE OF THE FE K$\alpha$ LINE EMISSION}
\label{properties}

{ \tablelinemeans summarizes various mean quantities from the
\fekalfa line measurements, calculated in two different ways.
In the first method we used the measurements from individual
observations and in the second method we used measurements
that are representative of properties {\it per source}.
For the latter, in most cases this utilized measurements from
spectra averaged over multiple observations where relevant,
except for 
NGC~526, PG~0834, 3C~273, IRAS~13349$+$2438 and 3C~382.
For these five sources the \fekalfa line was significantly detected
in only one observation and combining observations led to looser
constraints on the \fekalfa line parameters, as previously explained
in \S\ref{hegspec}. Thus we used only the one observation for
these five sources that showed the best detection of the
\fekalfa line. This may bias the results because we do not know
if non-detections of the \fekalfa are due to variability. We
caution that any sample properties derived using our results
should take account of such possible biases. We also caution that
the \chandra grating sample is subject to very peculiar and unquantifiable
selection effects because of the restrictions on the kind of
sources that are suitable for observations with gratings (or
more precisely, which sources proposal review panels judge to be suitable
for observations with gratings). Thus, the \chandra grating AGN archive
is not suitable for unbiased population studies. The principal purpose
of the present work is to systematically quantify the spectral
parameters from the data.}

\subsection{LINE CENTROID ENERGY}
\label{centroidenergy}

{From our analysis, we were able to measure the \fekalfa line
centroid energy in 32 out of 36 unique
sources for at least one spectrum (see \tablesigfixed and \tablesigfreep).
\tablelinemeans summarizes four different weighted mean line centroid
energies. One pair of measurements was derived from individual
observations and the other pair was derived from {\it per source}
measurements (as described above).
Each mean centroid energy was derived from spectral-fitting
results in which the intrinsic line width was fixed 
(\tablesigfixedp), and from results in which the intrinsic line width was
not fixed in all the spectra (\tablesigfreep).
Here, and hereafter, for the calculation of the weighted mean of
any quantity with asymmetric errors, we
simply assumed symmetric errors, using the largest 68\% confidence
error from spectral fitting.
For the line centroid values, 68 out of 82
spectra contributed to
the ``per observation'' means, and 32 sources contributed to the
``per source'' means. It can be seen from \tablelinemeans that
all four mean line centroid energies are within  $-12$~eV and $+3$~eV
of 6.400~keV (including statistical bounds).}

{ Note that the statistical errors on the mean
centroid energies are 1~eV or
better but they may be biased by the brightest sources and
largest exposure times. A
more useful measure of the dispersion in the line 
energies may be gleaned from
examining the distribution of energies.
\figecenhist shows histograms of the \fekalfa line
centroid energy. Again, we show four histograms:
\figecenhistp (a) and \figecenhistp (c) pertain to ``per observation''
results and \figecenhistp (b) and \figecenhistp (d)
pertain to ``per source'' results.
\figecenhistp (a) and \figecenhistp (b) pertain to 
\fekalfa line centroid energies measured with the intrinsic
line width fixed (\tablesigfixedp), and
\figecenhistp (c) and \figecenhistp (d)
pertain to results obtained when the intrinsic line width was not 
fixed in all the observations (\tablesigfreep).
The dashed and dotted lines correspond to histograms
made from the 68\% confidence
lower and upper limits on the line centroid energy respectively. 
The fits in which the \fekalfa line width was fixed
gave line centroid energies that are more reliable indicators
of the peak line energy because
the fits with the line width free are more prone to the
centroid energy being affected by the shape of the line profile.
All panels show that the histograms are not Gaussian but  
sharply peaked at $\sim 6.4$~keV. There is not a significant
difference between the ``per observation'' and ``per source'' distributions,
within the statistical errors.
For the ``per observation'' fits with the \fekalfa line width fixed,
we found that $\sim80\%$ of the best-fitting
line centroid energies lie in the
range  6.38--6.43~keV, a spread of only 50~eV.
Another way of expressing our results is that if we take
the highest signal-to-noise measurement for each source
(i.e. utilizing results from the summed spectra only, for
sources with multiple observations), we find that
21 out of 32 sources ($\sim 66\%$) have 
68\% confidence statistical bounds on the line centroid energy 
that lie entirely in the
range 6.38--6.43~keV. A similar procedure also shows that
30 out of 32 sources ($\sim 94\%$) have 68\% confidence
bounds on the line centroid energy that lie entirely in the
range 6.35--6.47~keV. We note that we might have expected
to observe additional peaks in the centroid energy distribution
blueward of 6.4~keV due to highly ionized Fe. 
Although such emission lines have been detected
in HEG data (e.g. NGC~7314, Yaqoob \etal 2003; NGC~7213, Bianchi
\etal 2008), the HEG
effective area is already very small at 6.4~keV (only $\sim 20 
\ \rm cm^{-2}$) and drops very rapidly at higher energies.
Higher throughput detectors such as those aboard \xmm or \suzaku
are more suitable for investigating the frequency of occurence    
of highly ionized Fe emission lines.}

We now examine those measurements that deviate 
significantly from the 6.4~keV peak of the \fekalfa
line centroid energy distribution in \figecenhistp.
For 3C~273 and 4C~74.26, we obtained 
\fekalfa line centroid energies lower than 
those for the bulk of the measurements, and we note that
even the 90\% confidence upper limits were less
than 6.4~keV for the fits in which the line width was
fixed (see \tablesigfixedp). For the fits in
which the line width was free, the corresponding
upper limits were 6.49~keV and
6.39~keV for 3C~273 and 4C~74.26 respectively (see \tablesigfreep).
However, for these two sources, 
the detection of the \fekalfa line was
marginal: $C$ decreased by
less than 9.3 when a narrow Gaussian was added to a power-law 
continuum-only model.
Thus, the lines were detected with only 99\% confidence or less.
Such low centroid energies are not unphysical. For example
they could be affected by
gravitational redshifts.
We note that a weak broad \fekalfa line has 
been detected in 3C 273 
by $XMM-Newton$ (Page et al. 2004a) and the 99\% confidence 
contour does not rule a line 
with a centroid energy in the range $\sim$ 6.2-6.3 keV. 
Interestingly, a narrow component of the \fekalfa line
at 6.2~keV, in addition to an \fekalfa line 
at $\sim 6.4$~keV, has been detected in 4C 74.26 by $XMM-Newton$ 
(Ballantyne \& Fabian 2005). 

For PG~0844$+$346, we obtained a centroid energy
for the \fekalfa line consistent with 6.4~keV from the
fits with the line width fixed (see \tablesigfixedp).
However, allowing the line width to float
gave a centroid energy of $\sim 6.6$~keV, with a 90\%
confidence lower limit of 6.42~keV (see \tablesigfreep).
The reason for this apparently discrepant behavior
is clear from the spectrum of PG~0844$+$346 in \figindspecp.
The spectrum shows an emission line centered at $\sim 6.4$~keV
and this is picked up in the fits for which the line width was
fixed (the measured energy was 6.364$^{+0.007}_{-0.009}$~keV).  
The spectrum also shows two additional peaks at higher energies,
albeit with a low signal-to-noise ratio. 
Allowing the single-line Gaussian model
width to be free in the fits then causes the line component
to model all three narrow lines by broadening the Gaussian.

In summary, we measured the centroid energy
of the \fekalfa emission line in 32 out of 36 sources.
In 30 out of the 32 sources the line
centroid energy lies in the range 6.35--6.47~keV, inclusive
of the 68\% confidence statistical errors.
We note that ionization states
less than Fe~{\sc xvii} 
correspond to \fekalfa line energies lees than 6.43~keV
(e.g. Palmeri \etal 2003; Mendoza \etal 2004).
When individual sources amongst the 30 are considered,
the line centroid energy can constrain the ionization state
to be much lower than Fe~{\sc xvii} in some cases. 
In the remaining 2 sources, 3C~273 and 4C~74.26,
the line centroid energy, including statistical errors,
appears to be lower than 6.4~keV,
but the detection is marginal in these two sources.

\subsection{LINE EQUIVALENT WIDTH}
\label{lineew}

{ From the spectral-fitting results for which the
\fekalfa line intrinsic width was fixed (\tablesigfixedp),
33 of the 36 sources have at least one spectrum from
which we could measure the EW with a non-zero 90\% confidence
lower limit and a finite upper limit. 
The three sources for which only upper limits on the EW could be
obtained were Mkn~705, PDS~456, and IRAS~18325$-$5926.
In total, 70 out of 82 of the individual observations in \tablesigfixed yielded 
bounded lower and upper limits on the EW.
We calculated weighted mean EW values in four different
ways (as we did for the line centroid energy in \S\ref{centroidenergy}):
i.e. from ``per observation'' and ``per source'' values, each set obtained
with the intrinsic line width fixed in all spectral fits
(\tablesigfixedp) and with the intrinsic line width free
in some of the spectra (\tablesigfreep).
The resulting mean EW values are summarized in \tablelinemeansp.
It can be seen that the mean \fekalfa EW is somewhat sensitive to how
it is calculated, ranging from $42$~eV to 70~eV, with a statistical
error of 4~eV or less. We note, however that a value of $\sim 40$~eV
could be interpreted as a fairly robust sample lower limit on the
EW of any unresolved core of the \fekalfa line. }

{
\figewhist shows histograms of the \fekalfa line
EW, again constructed in four different ways. 
\figewhista and \figewhistc pertain to ``per observation''
results and \figewhistb and \figewhistd
pertain to ``per source'' results.
\figewhista and \figewhistb pertain to
\fekalfa line EWs measured with the intrinsic
line width fixed (\tablesigfixedp), and
\figewhistc and \figewhistd 
pertain to results obtained when the intrinsic line width was not
fixed in all the observations (\tablesigfreep).}

{
The dashed and dotted histograms in \figewhist correspond to
the distributions of the 68\% confidence 
lower and upper limits on EW, respectively. 
The shaded histograms in both panels are the 68\% upper limits on
the EW for the 12 observations in  
which the EW could not be measured. 
For the ``per observation'' and intrinsic line width fixed results,
\tablesigfixed and \figewhista
show that the maximum best-fitting EW of the \fekalfa line core 
is $162$ eV, and $>$90\% of the measurements have a 
best-fitting EW less than 100 eV. 
We also found that 79\% of the measurements have a 
68\% confidence upper limit on the EW of less than 100~eV (23 unique sources).
Including the results from the fits with the \fekalfa line
width free, we found that
$\sim70\%$ of the ``per observation'' measurements have a 
best-fitting equivalent EW less than 100~eV (\figewhistcp).
Within the statistical errors, the histograms obtained from the
``per observation'' results are not significantly different to the
corresponding ``per source'' histograms.
Note that in \figewhistc and \figewhistdp,
we do not show the measurement for PG~0844$+$346, as its 
EW is artificially high ($\sim$ 600 eV) because
it is not a true measure of the EW of the emission
line at $\sim 6.4$~keV (see \S\ref{hegspec}).}

From a theoretical point of view,
the \fekalfa line EW depends on a number of factors, including
geometry, orientation, column density, and
covering factor of the
line-emitting matter distribution, as well as element abundances.
Time delays between variations in the continuum level
and the \fekalfa line flux also affect the EW measured
during a given observation. The sample EW results should
therefore be interpreted in terms of a particular geometry.
The relatively small dispersion of the EW distribution that we measure
from the HEG data translates into a small dispersion
in the parameters mentioned above, but it is difficult
to uniquely constrain these parameters from the EW distribution
due to degeneracy. In the context of the toroidal X-ray reprocessor
model of MY09, which subtends a solid angle
of $2\pi$ at the X-ray source, the measured EW distribution is
consistent with the MY09 model if the mean of
the EW distribution corresponds to column densities greater than
$\sim 2 \times 10^{23}  \rm \ cm^{-2}$ (see MY09). 
This column density does not refer to the line-of-sight
value, but rather to the angle-average over all incident X-ray continuum
radiation. Comparison
of the HEG results with the toroidal reprocessor models of 
Ghisellini, Haardt, \& Matt (1994) and 
Ikeda, Awaki, \& Terashima (2009) leads to similar
conclusions. The upper bound on the column density is not
constrained because, for situations in which the \fekalfa
line is observed for lines of sight that intercept 
a column density $<10^{23}  \rm \ cm^{-2}$, the 
EW attains a maximum for an angle-averaged column density of
$\sim 10^{24}  \rm \ cm^{-2}$, above which the EW decreases again
as the line-emitter becomes Compton-thick (e.g. se MY09).
Increasing the covering factor of the reprocessor can increase
the EW of the \fekalfa line observed in reflection but 
there is a trade-off because as the covering factor approaches
unity, the projected area of the reflection region decreases
and emission-line photons are more prone to being impeded from
escaping the medium. Ikeda \etal (2009) found that the EW
is greatest for covering factors factors of $\sim 0.7-0.9$ but
does not exceed $\sim 180$~eV in their model for cosmic abundances
and a power-law photon index of $1.9$.
In principle, the shape and relative magnitude 
of the \fekalfa line Compton shoulder could determine whether
the reprocessor is Compton-thin or Compton-thick but this
is challenging due to the limited signal-to-noise ratio of the data and
also requires more sophisticated modeling. Such an investigation
will be reported in future work. So far, all analyses with
respect to the Compton shoulder and HEG AGN data have employed {\it ad-hoc} 
models (e.g. Kaspi \etal 2002; Yaqoob \etal 2005)
so they do not yield a meaningful physical interpretation. \\

\subsection{INTRINSIC LINE WIDTH}
\label{linewidth}

The location of the medium responsible for
the core of the \fekalfa emission line
can potentially be constrained by the measurements of the line intrinsic width.
The weighted mean FWHM of the \fekalfa line cores for the 53 
individual data sets (27 unique sources) for which it could be measured
(\tablesigfreep), is $2060 \pm 230 \ \rm 
km \ s^{-1}$. This includes the two sources (Mrk~290 and 4C~74.26) for
which the \fekalfa line FWHM could only be constrained
from the summed spectra. { We also calculated the weighted mean
FWHM from ``per source'' measurements and found a
similar value of $2200 \pm 220 \ \rm km \ s^{-1}$ (see \tablelinemeansp).}

In \tablesigfree are values of the \hbeta FWHM
compiled from the literature. Comparing the \fekalfa line
FWHM with that of the \hbeta line can potentially
give a direct indication of the location of the \fekalfa
line-emitting region relative to the optical broad-line region.
A direct comparison of the \fekalfa line width with optical BLR
line widths was not attempted in YP04 because the number of
sources with sufficiently high quality \fekalfa line-width measurements
was too small. Nevertheless, Nandra (2006) using the YP04 results,
supplemented by a few other HEG measurements from the literature,
examined the
relation between the FWHM of \fekalfa and \hbetap. The results
were ambiguous, the data allowing for an origin of the \fekalfa line
anywhere from the BLR to parsec-scale distances from the putative central
black hole. Moreover, some of the HEG measurements for the \fekalfa
line FWHM compiled from the literature were problematic. For example,
for MR~2251$-$178, { Gibson \etal (2005) 
reported an 
upper limit on the \fekalfa line FWHM of $1530 \ \rm km \ s^{-1}$
and Nandra \etal (2006) erroneously quoted and used 
as $650 \ \rm km \ s^{-1}$ (Gibson \etal 2005 reported a $\sigma$ of
$650 \ \rm km \ s^{-1}$, not FWHM). }
In our uniform analysis, we found that
the HEG data for MR~2251$-$178 were so poor that a meaningful
upper limit on the \fekalfa FWHM
cannot even be measured and therefore we have reported only results
for a fit with the line width fixed at well below the HEG resolution 
(\tablesigfixedp).

A notable example for which a meaningful comparison between
the \fekalfa and \hbeta line widths has been reported using
HEG data is NGC~7213 (Bianchi \etal 2008). In this case
the FWHM of both lines are consistent with each other 
($\sim 2500 \ \rm km \ s^{-1}$), implying an origin of the
\fekalfa line in the BLR for NGC~7213. Our sample includes
NGC~7213 and our analysis (see \tablesigfreep)
confirms the results of Bianchi \etal (2008).
Utilizing all of the results from our uniform analysis of the
HEG sample for which the \fekalfa line width could be at least loosely
constrained
(\tablesigfreep), we have plotted in \figfwhmvshbeta the \fekalfa
line FWHM against the \hbeta FWHM. The dashed line corresponds
to the two line widths being equal. The statistical errors shown
correspond to 68\% confidence. 
We have distinguished 12 sources in \figfwhmvshbeta by
empty circles (as opposed to filled circles) that provide
the very best statistical constraints on the \fekalfa line FWHM in our
sample. The next best measurement of the FWHM would 
be NGC~985, but we note that its 99\% two-parameter confidence contour
for \fekalfa line flux versus FWHM did not close before the
Gaussian component began to model the continuum.
In \figfwhmvshbetap, points that lie above
the dashed line at some level of confidence mean that an origin
in the BLR of at
least part of the \fekalfa line is not ruled out,
but contributions from further out than the BLR
are not ruled out either (at the appropriate level
of confidence). A larger FWHM for the \fekalfa line compared
to the \hbeta line could either mean a genuine contribution
to the \fekalfa line from matter closer to the central
black hole than the BLR, or it could mean that there is
a contribution from an unresolved Compton shoulder or from
part of a broader disk line.
Points that lie {\it below} the dashed line in \figfwhmvshbeta at some
level of confidence place stronger constraints on the
origin of the narrow \fekalfa line because in that case, whatever
physical sources of broadening are affecting
the \fekalfa line, it must originate in a region that lies
further from the central black hole than BLR.

Standard tests for assessing the significance of any possible
correlation between FWHM(\fekalfa) and FWHM(\hbeta),
such as the Spearman Rank correlation coefficient, are 
problematic because they do not take account of
measurement errors on FWHM(\fekalfa), which can be large.
{ Assessing the effect of measurement errors 
on such correlation coefficients properly
requires extensive and realistic simulations of the
data {\it and} the spectral-fitting process.}
Instead, we used the $\chi^2$ statistic
to fit a straight line to the FWHM values of 
the \fekalfa and \hbeta lines.
Although we are forced to assume a relationship between the two
quantities, any correlation would still manifest itself.
In the fitting we explicitly took
took into account the statistical errors on the \fekalfa 
line widths, using the average of the 68\% confidence upper and lower errors. 
We found that FWHM(\fekalfa)=(0.04$\pm0.13)\times$ FWHM(\hbeta)+(2130$\pm550$), 
with $\chi^2=17.1$ for 21 degrees of freedom. The reduced $\chi^{2}$
value $<1$ then means that indeed a more complicated model is not
warranted. More importantly,
we see that even the $1\sigma$ errors on the slope include a
slope value of zero (corresponding to the case that all the
FWHM values are consistent with a constant, independent of the
FWHM of the \hbeta line). 
Therefore, we find no evidence of a correlation 
between the \fekalfa and \hbeta line widths, consistent with 
the conclusion of Nandra \etal (2006).
However, Nandra \etal (2006) interpreted
the lack of a correlations in terms of the narrow
\fekalfa line not originating in the BLR, but we now know that
in some cases this is not true (e.g. NGC~7213, Bianchi \etal 2008).
We shall see
below that our results in fact show that the location of
the \fekalfa line emitter relative to the BLR 
appears to be genuinely different 
from source to source.

From our spectral fits to the subset of
HEG data with the \fekalfa line width free (\tablesigfreep)
we constructed joint 68\%, 90\%, 99\% confidence contours of
the \fekalfa line EW versus the {\it ratio}
of the \fekalfa FWHM to the \hbeta FWHM. These are shown in
\figifevsfwhm and a variety of behavior is displayed.
We found cases in which this FWHM ratio was, at
the two-parameter 99\% confidence level, was
less than 1 (NGC~3783, NGC~4151 and NGC~5548), greater than 1
(MCG~$-$6-30-15), or consistent with 1 
(3C~120, NGC~2110, MCG~$-$5-23-16, NGC~3516, NGC~5506, Mrk~509,
NGC~7213, and NGC~7469).
Thus, it appears that the
location of the \fekalfa line relative to the location of the \hbeta
line-emitting region may be different
from source to source. For our limited-sized
sample the \fekalfa line-emitting region size could be up to 
a factor $\sim 5$ larger than the \hbeta line-emitting region
(NGC~4151 -- see \figifevsfwhmp). 
We note that the putative parsec-scale obscuring torus that is required by
AGN unification schemes, and that has always been a strong contender
for any \fekalfa line emission beyond the BLR, may be smaller
than traditionally thought. In particular,
Gaskell, Goosmann, \& Klimek (2008) argue that there is considerable 
observational evidence that the BLR itself has a toroidal structure, 
and that there may be no distinct boundary between the
BLR and the classical parsec-scale torus. 
Our results from the \chandra HEG data do not conflict with such
a scenario.

From joint confidence contours of \fekalfa line intensity 
versus FWHM we can determine
whether the line is resolved from a given data set
if the contour (at some level of confidence) does not
cross the $\rm FWHM=0$ axis.
We found that at 99\% confidence
(two parameters), the \chandra HEG resolves the narrow
component of the \fekalfa emission in 15 sources, namely, F9, NGC~2110, 
MCG~$-$5-23-16, NGC~3516, NGC~3783, NGC~4051, NGC~4151, 
MCG~$-$6-30-15, IRAS~13349$+$2438, IC~4329A, Mrk~279, NGC~5548,   
E1821$+$643, NGC~7469 and NGC~7213. 
We do not include PG~0834$+$346 here, since the single-Gaussian fit 
with the line width free does not 
pick up the narrow component at $\sim$ 6.4 keV (see \S\ref{hegspec}).  
We caution that in general an emission line that is resolved by the
HEG may indicate complexity as opposed to a simple, single emission line.

\subsection{LINE FLUX}
\label{lineflux}

If the \fekalfa line originates in a matter distribution whose
light-crossing time is much greater than the typical timescale
of variability of the X-ray continuum, the variability 
of the line flux will be suppressed. The line flux may then
even be constant (within statistical errors) and
correspond to some historically-averaged continuum level.
The sources in our HEG sample that have multiple observations
enable us to investigate the time-dependence of the \fekalfa line
flux. The spectral resolution of the HEG currently allows the best isolation
of the narrow \fekalfa line for the largest sample compared 
to previous studies. In \figspectot we showed the \fekalfa line
intensity versus centroid energy 99\%, two-parameter confidence contours for
each source that has multiple observations. The contours for
NGC~3783 were shown in Yaqoob \etal (2005) and are not shown again
in \figspectotp. In no source did we find evidence for
variability of the \fekalfa line flux at 99\% confidence or greater.
However, it is important to note that the 99\% confidence
regions in some cases cover a large range in line flux due
to limited signal-to-noise ratio. However, we can say that in our HEG
sample, the data are consistent with no variability of the \fekalfa
line but more sensitive instrumentation is required to
reduce the statistical errors.
 
\section{X-RAY BALDWIN EFFECT}
\label{baldwineffect}

The so-called X-ray Baldwin effect, a possible anti-correlation
between the \fekalfa line EW and X-ray luminosity, has been discussed
at length in the literature (e.g. Iwasawa \& Taniguchi 1993; Nandra \etal 1997;
Page \etal 2004b; Jiang, Wang, \& Wang 2006; 
Bianchi \etal 2007; Winter \etal 2009).
These studies have found some evidence for an X-ray Baldwin effect
albeit with significant scatter, but
the latter two studies have found that the \fekalfa line EW
appears to be more strongly anti-correlated with the ratio
of X-ray luminosity to Eddington luminosity ($L_{x}/L_{\rm Edd}$,
a proxy for the accretion
rate). However, Winter \etal (2009) found that the X-ray Baldwin
effect was only significant if the EW and \lxledd values
were binned, and the
formal significance of the anti-correlation depended strongly on the
details of the binning procedure.  
Except for some \hetg data used by 
Jiang et al. (2006), {\it all} other studies
of the X-ray Baldwin effect to date have been based on data
that has a spectral resolution of $\sim 7000 \ \rm km \ s^{-1}$~FWHM
or worse. Therefore, it is not clear whether the
\fekalfa line parameters in these studies correspond to contributions
from line emission blended from completely different origins
(e.g. distant-matter and accretion-disk components).
Using our sample that consists {\it only}
of HEG data, we can investigate the X-ray Baldwin
effect with a spectral resolution in the Fe~K band that is
nearly four times better than in previous studies, and therefore
provide the best
isolation of the narrow core that is possible
with current instrumentation. 

{ For this purpose we used our
spectral-fitting results obtained
with the \fekalfa line width fixed
at 1~eV, well below the HEG resolution, in order to obtain a
uniform set of \fekalfa line EW measurements for the largest 
number of sources (see \tablesigfixedp). 
We examined correlations using both the ``per observation''
results and the ``per source'' results.
Measurements for
the latter were derived from only one spectrum per source,
which in some cases was the average spectrum, as described 
in \S\ref{properties}.
These values of
EW are plotted against
$L_{x}$ in \figbaldwina and \figbaldwincp, and against
\lxledd in \figbaldwinb and \figbaldwindp.
Note that in \figbaldwin we have shown all EW measurements
whether or not they are only upper limits, even though
upper limits will not be used in the quantitative analysis.}
The Eddington luminosity, $L_{\rm Edd}$, is computed from 
$M_{\rm BH}\times1.3\times10^{38}$ erg s$^{-1}$, where
$M_{\rm BH}$ is the mass of the central black hole. 
Values of $M_{\rm BH}$ are from Zhou \& Wang (2005), Bianchi 
et al. (2007), and Wang et al. (2009). 
As a proxy for the accretion rate we 
use the ratio of ${L_{\rm 2-10 keV}/L_{\rm Edd}}$
(see e.g. Vasudevan \& Fabian 2009 for the correspondence
between and X-ray luminosity and bolometric luminosity).
{ We were not able
to find reliable mass estimates for Mrk~705 and 
IRAS~18325$-$5926 so these sources were excluded from
any analyses involving $L_{\rm Edd}$.}
The statistical errors shown in \figbaldwin
are 68\% confidence for two free (Gaussian) parameters.
It can be seen that, despite better isolation of the \fekalfa line core,
there is still significant scatter in the diagrams.

In order to formally assess the significance of any correlation,
standard methods that do not take account of
the statistical errors on the EW,
such as the Spearman Rank correlation coefficient { are problematic}.
This is because, in the type of analysis presented here,
and in previous works on the X-ray Baldwin effect,
the actual best-fitting values of EW are not in themselves
meaningful. It is the statistical errors on the EW that are
the important quantities.
{ Assessing the effect of measurement errors 
on such correlation coefficients properly
requires extensive and realistic simulations of the
data {\it and} the spectral-fitting process.}
On the other hand, the $\chi^{2}$ statistic does take account
of the statistical errors on the EW. Although we are forced
to assume a form of the relationship between the EW and
$L_{x}$ if we use
$\chi^{2}$, it can be seen from \figbaldwin that the quality of
the data do not support constraining a more complex relationship.
We therefore fitted a straight line to $\log{EW}$ versus $\log{L_{x}}$
using the $\chi^{2}$ fit statistic
(i.e. a power-law function for EW versus $L_{x}$). 
Data points that only had upper limits on the EW
were {\it not} included. In the fitting we took
into account the statistical errors on the 
EW, using the average of the 68\% confidence upper and lower errors.

{ The results of the $\chi^{2}$ analysis are shown in \tablebaldwinp.
For each of the four cases (``per observation'', ``per source'', and
EW versus $L_{x}$ or $L_{x}/L_{\rm Edd}$) we show the best-fitting
value of $\chi^{2}$, the intercept and slope of the best-fitting line,
as well as the
the 68\% confidence and 99\% confidence one-parameter errors on the slope.
The latter error bounds were determined by varying the slope,
whilst allowing the intercept to float, and
determining the bounds on the slope for $\Delta\chi^{2}=0.989$ and
6.635 for 68\% and 99\% confidence respectively.
If the EW is indeed anti-correlated with either $L_{x}$ or $L_{x}/L_{\rm Edd}$
we would expect that the slope of the line is significantly different
from zero. Therefore, in \tablebaldwin we
also show values of
$\Delta \chi^{2}$ obtained when the slope is forced to be
zero, as well as the corresponding significance that the slope
is non-zero.
We found that the ``per observation'' results gave a stronger anti-correlation
than the ``per source'' results, for both the EW versus $L_{x}$ and EW
versus $L_{x}/L_{\rm Edd}$ relations. Quantitatively, the ``per observation''
results show a significance of $6.05-6.27\sigma$ for a non-zero slope,
as opposed to $\sim 3.08-3.24\sigma$ for the ``per source'' results. 
The best-fitting slopes for the latter are about half of the
corresponding values of the ``per observation'' results.
We caution that the absolute significance values should not
be interpreted literally since we do not know the form of the 
functional relationship between EW and $L_{x}$.
\tablebaldwin also shows that there is no significant difference
in the $\chi^{2}$ analysis results on whether we examine the
relation of EW between $L_{x}$ or $L_{x}/L_{\rm Edd}$, and that is
true whether we consider the ``per observation'' or ``per source'' results.}
{ Both our
``per observation'' and ``per source'' results for the 
slope of the EW versus $L_{x}$ relation 
are formally consistent, within the uncertainties, with 
that found by Page et al. (2004b) who 
reported $EW~\propto~L^{-0.17\pm0.08}$).   
In addition, our results for the slope of the 
relation between EW and $L_{x}/L_{\rm Edd}$ is formally
consistent with that 
obtained by Bianchi et al. (2007) 
(${\rm EW}~\propto~(L_{\rm bol}/L_{\rm Edd})^{-0.19\pm0.05}$).
We note that the latter study of Bianchi \etal (2007) excluded
sources with high radio-loudness and still found a significant
Baldwin effect.}

{ Our results seem to confirm the X-ray Baldwin effect.
There are several factors that could produce an anti-correlation of
the EW of the \fekalfa line and the intrinsic 
X-ray continuum luminosity. A decrease of
covering factor and/or the column density of line-emitting 
with increasing X-ray continuum luminosity likely are the 
most important factors. Another possibility is
that the line-emitting material becomes more and more ionized as
the X-ray luminosity increases, leaving less low-ionization material to 
produce the \fekalfa line at $\sim 6.4$~keV.
Unfortunately the data cannot yet distinguish between these scenarios.
A complete understand of the Baldwin effect should also 
take into consideration the fact that
the \fekalfa line EW in individual sources can vary by
more than a factor of two (if the line intensity does not respond
to large-amplitude continuum variations), although simulations based
on the simplest assumptions yield an anti-correlation between EW and continuum 
luminosity weaker than observed ones and with large scattering 
($EW~\propto~L^{-0.05\pm0.05}$, Jiang et al. 2006).
}

\section{SUMMARY}
\label{summary}

We have presented an empirical and uniform analysis of the
narrow core of the \fekalfa emission line in a sample of 
82 observations of 36 AGNs
with low to moderately low X-ray absorption ($N_{H}<10^{23} \ \rm cm^{-2}$),
using \chandra HEG data. The \fekalfa line was detected
in 33 sources, and its centroid energy was measured in 32 sources
(68 observations). The distribution in the centroid energy is
strongly peaked around $\sim 6.4$~keV, with over 80\% of the
measurements lying in the range 6.38--6.43~keV. Including the
statistical errors and utilizing the best measurements for
each source, the line centroid energy lies entirely
in the range 6.35--6.47~keV for 30 out of 32 sources. 
Thus we confirm, for the largest sample of AGN observed with
such a high spectral resolution (FWHM~$\sim 1860
\ \rm km \ s^{-1}$ at 6.4~keV), the ubiquity of the 
narrow core of \fekalfa line, and its preferred origin in
cool, neutral or only mildly-ionized matter.

The equivalent width (EW) of the core of the \fekalfa line
was constrained in 70 out of 82 observations, with only upper
limits obtained from the remaining 12 spectra.
The weighted mean EW was $53 \pm 3$~eV, and $\sim 70\%$ of the
individual measurements had a 68\% confidence upper limit on
the EW of less than 100~eV. { Similar
results were obtained when considering the EW distribution
by source, although the weighted mean was somewhat
higher from measurements that allowed the intrinsic
line width to be free ($70\pm4$~eV).} The EW distribution can be
produced by both Compton-thin and Compton-thick matter distributions
and a more detailed analysis with a physical model is required
to distinguish between the two scenarios.
We also presented measurements of the flux of the core of the
\fekalfa line and found that for sources that had multiple
observations, there was no case in which the line flux
varied between observations, within the statistical errors.

The intrinsic width of the core of the \fekalfa line was measured
for 27 sources (53 observations) and we obtained a weighted mean
value of FWHM~$=2060 \pm 230 \ \rm km \ s^{-1}$ {
(or $=2200 \pm 220 \ \rm km \ s^{-1}$ when considering
measurements by source, not by observation)}. Of the 27 sources,
12 yielded 99\% confidence, two-parameter contours of line flux versus
FWHM that were good enough to investigate the relation between
the width of the \fekalfa line and the width of the H$\beta$ line
(or Br$\alpha$ for one of the sources). We found that the ratio
of the X-ray to optical line width varies from source to source.
The 99\% confidence, two-parameter upper limit lies 
in the range $\sim 0.5-4$ for the 12 sources. This means
that contributions to the flux of the core of the \fekalfa line
are allowed down to a factor $\sim 0.7-2$ times the radius
of the optical BLR. The upper limit on the size of the X-ray line
emitter is not constrained because line flux contributions from
large, parsec-scale distances could be unresolved by the HEG.
We note that our results are suggestive of the fact that
the location of the X-ray line-emitting region relative to the
BLR may actually be different in different sources. 
These conclusions are subject to the caveat that derivation
of the true velocity width of the \fekalfa line core requires
a proper physical model, such as that of MY09, 
that includes a possible Compton shoulder.
This will be the subject of future work. However, we note that
such an analysis can only reduce the derived velocity widths of the
\fekalfa lines. 
Finally, having isolated the narrow core of the \fekalfa line
with the best available spectral resolution we 
confirm the
anti-correlation (albeit with a large scatter)
between the line EW and X-ray luminosity, $L_{x}$
(the X-ray Baldwin effect),
and between the line EW and $L_{x}/L_{\rm Edd}$. 

We thank the referee for his/her useful comments. 
Partial support for this work was provided by NASA through \chandra Award
AR8-9012X, issued by the Chandra X-ray Observatory Center,
which is operated by the Smithsonian Astrophysical Observatory for and
on behalf of the NASA under contract NAS8-39073.
X.W.S. and J.X.W. acknowledge support from Chinese National Science Foundation (Grant No. 10825312, 10773010), and Knowledge Innovation Program of CAS (Grant No. KJCX2-YW-T05).
This research
made use of the HEASARC online data archive services, supported
by NASA/GSFC. This research has made use of the NASA/IPAC Extragalactic Database
(NED) which is operated by the Jet Propulsion Laboratory, California Institute
of Technology, under contract with NASA.
The authors are grateful to the \chandra 
instrument and operations teams for making these observations
possible.

\section{APPENDIX: NOTES ON INDIVIDUAL SOURCES}
\label{appendix}

In this section we give, for each source in our sample,
particular additional details of the analysis
and/or results where necessary. We also summarize briefly
any previously-published HEG results in the Fe~K band
that are based on the same data. Our intention is not to
review observations by other instruments.

{\it F9}. Chandra HEG results were reported in YP04
and the new analysis is consistent with the previous results.
Note that the very large upper limit on the intrinsic \fekalfa line
width (\tablesigfreep) is unphysical since such a broad Gaussian
component is clearly modeling the underlying spectrum
(see discussion in YP04).
The HEG data show marginal evidence of an emission line at $\sim 6.9$~keV.

{\it NGC~526a}. No results from either of the two observations
have been previously published. The \fekalfa line is detected in only
one of the observations, and the detection is marginal. Consequently,
the FWHM of the line could not be constrained.

{\it Mrk 590}.
Results from the
\chandra HEG data have been presented by Longinotti et al. (2007),  
who reported the detection of a narrow \fekalfa line with 
\fekecenp$=6.40^{+0.04}_{-0.03}$~keV, 
\feksigmap$=47^{+58}_{-24}$~eV, and ${\rm EW}=160^{+118}_{-78}$~eV. 
Our best-fitting \fekalfa line parameters (\tablesigfixedp) are in 
good agreement with those measured by Longinotti et al. (2007). 
\figindspec shows that the large 99\% confidence region for \fekflux versus
\fekecen indicates that the fits in which the line width was free
do not provide a reliable measure of the intrinsic line width.
 
{\it NGC~985}. Although Krongold \etal (2005) reported results
from the \chandra \hetg observation, they combined HEG and MEG data
and did not report results on the \fekalfa line emission.
 
{\it ESO~198$-$G24}. No results from any of the two \chandra \hetg
observations have been previously reported. We obtained a significant
detection of the narrow \fekalfa line from only one of the observations
(\tablesigfixed and \tablesigfreep).

{\it 3C~120}. Results 
on the \fekalfa line from the \chandra \hetg observation of this
source have been reported in YP04 and the new results presented here
are consistent with the previous ones. The HEG data show a
marginal detection of an emission line at $\sim 6.9$~keV (YP04).

{\it NGC 2110}.
The results from the
four \chandra \hetg observations were presented by Evans et al. (2007), 
who measured the narrow \fekalfa line parameters 
\fekecen~$=6.397\pm0.007$~keV,
and ${\rm EW}=81^{+27}_{-30}$ eV, consistent with our results. 
Note that in the second observation the line width could not
be constrained so the \fekflux versus \fekecen 99\% confidence contour
for that observation in \figspectot was constructed with
the line width fixed at 1~eV (dot-dashed line).

{\it PG~0844$+$349}. No results from any of the three \chandra \hetg
observations of this source have been reported previously.
The detection of the \fekalfa line
at $\sim 6.4$~keV is marginal, and there is also marginal evidence
of emission lines due to He-like and H-like Fe. When fitted
with a single-Gaussian model, the presence of three narrow emission
lines causes the Gaussian intrinsic width to become large as it
tries to account for all three lines. Therefore, the most reliable
values of \fekecen and \fekflux are those obtained from fits in 
which the line width was fixed. 

{\it MCG~$-$5-23-16}. This source was observed
by the \chandra \hetg on three occasions.
Results from the first observation have
been presented by Balestra et al. (2004), who found,
from single-Gaussian fits
to the narrow \fekalfa line,  \fekecenp$=6.38\pm0.02$~keV, 
${\rm EW}=70\pm28$~eV, and FWHM $\leq$ 6500 km s$^{-1}$ 
(at 99\% confidence). Results from the remaining two 
observations were presented by Braito et al. (2007), who reported 
\fekalfa narrow-line parameters for the mean (time-averaged) spectrum of
\fekecenp$=6.41^{+0.02}_{-0.01}$~keV, and ${\rm EW}=61^{+17}_{-23}$~eV.
These correspond to the case when the line width was fixed at
a value less than the instrument resolution and Braito \etal (2007)
found that if the line width was allowed to float, the constraints
were sensitive to details of the continuum and relativistic disk-line model. 
Our results are consistent with previously published results; our simple
continuum model and omission of a broad relativistic line in the
fits means that our measurements of the line width should be 
interpreted as empirical indicators only.
Note that in the second observation the 99\% confidence contour of the 
\fekalfa line intensity versus energy was not closed 
when the line width was a free parameter. 
Thus, for this observation, we show in \figspectot the
99\% confidence contour for the line width fixed at 1~eV
(thin solid line).

{\it Mrk~705}. The signal-to-noise ratio of the data in this observation
was very poor. Previous results have been reported by Gallo \etal (2005)
who obtained an upper limit of 149~eV on the EW of an emission line
with a centroid energy fixed at 6.4~keV. This is consistent with our
analysis (\tablesigfixedp).

{\it NGC~3227}. Results from the \chandra \hetg observation of this
source have been reported in YP04 and the new results presented here
are consistent with the previous ones.

{\it NGC 3516}. There were eight \chandra \hetg observations of this source. 
Results from the first three 
observations were reported in YP04.
In the present paper we report on the analysis of five new observations 
that were performed in October 2006. Results from the same datasets 
have also been presented by Turner et al. (2008), who reported 
the detection of a narrow \fekalfa emission line with
\fekecenp~$=6.404\pm0.019$~keV, \feksigma~$=40^{+10}_{-15}$~eV,
and ${\rm EW} \sim 94$~eV (the statistical error was not given).
In addition, redshifted emission-line features
have been reported in some of the HEG data (Turner \etal 2002),
as well as H-like and He-like Fe emission and absorption features 
(Turner \etal 2008).
In the present paper we are concerned only with the \fekalfa 
emission line centered at $\sim 6.4$~keV and our
results are consistent with those of Turner \etal (2008). 
Due to the short exposure time of the last observation, 
the \fekalfa line was detected at
less than 99\% confidence (for two free Gaussian parameters). 
Thus, we do not show the contour of the line intensity versus energy 
in \figspectot for this observation.          

{\it NGC~3783}. Detailed results from the six \chandra \hetg
observations of this source have been presented by Yaqoob \etal (2005),
and Kaspi \etal (2001, 2002) and our re-analysis is consistent with
the previous results.

{\it NGC~4051}. Results from the \chandra \hetg observation of this
source have been reported by Collinge \etal (2001), who obtained
\fekecenp~$=6.41^{+0.01}_{-0.01}$~keV, ${\rm EW}=158^{+51}_{-47}$~eV,
and FWHM~$<2800 \rm \ km \ s^{-1}$ for the core of the narrow
\fekalfa line. Our analysis is consistent with the previous
results, except that when the line width was free in the
fits it becomes larger than the narrow core in the data (\tablesigfreep),
and this is consistent with the results reported in YP04.
Absorption features due to He-like and H-like Fe have also
been noted in the HEG data for NGC~4051 (Collinge \etal 2001; YP04).

{\it NGC~4151}. This source was observed five times with the
\chandra \hetgp. Results from the first observation have been
reported by Ogle \etal (2000) who obtained ${\rm EW}=160 \pm 20$~eV,
consistent with our measurement (\tablesigfixedp) and a
FWHM of $1800 \pm 200 \rm \ km \ s^{-1}$, also consistent with
our analysis (\tablesigfreep). The line centroid energy was not 
measured. Results for narrow \fekalfa line
parameters measured by the HEG for the remaining four observations
have not been previously published.

{\it Mrk~766}.
Results for the \chandra \hetg observation of this source
have been reported in YP04 and our re-analysis is consistent with the
previous results. The HEG data show marginal evidence of an
emission line at $\sim 6.9$~keV (YP04). 

{\it 3C~273}. A narrow \fekalfa line was detected in only one of
the 7 \chandra \hetg observations of this source. Measurements
of the \fekalfa line core from \chandra HEG data have not
been previously reported. 

{\it NGC~4593}.
Results for the \chandra \hetg observation of this source
have been reported in YP04 and our re-analysis is consistent with the
previous results. The HEG data show marginal evidence of an
emission line at $\sim 6.9$~keV (YP04).

{\it MCG~$-$6-30-15}.
This source was observed by the
\chandra \hetg five times. Results
from the first observation have been
presented by Lee et al. (2002) and YP04. 
The results for the other four observations were presented by Young 
et al. (2005), who reported narrow \fekalfa
emission-line parameters from the
time-averaged spectrum of { \fekecenp~$=6.393^{+0.106}_{-0.014}$~keV},
${\rm EW}= 18^{+11}_{-8}$~eV,
and a FWHM~$<4700 \rm km \ s^{-1}$. 
In the present analysis, only one out of four new observations had 
a significant detection of the narrow \fekalfa
emission line. From our empirical analysis we obtained
a larger EW and FWHM than Young \etal (2005). This could be
attributed to a contribution to the \fekalfa line
core from an underlying disk-line
component and/or the difference could be due to a complex continuum.
However, there is a large range of possible models but in our
analysis the simple empirical model is appropriate because
the results can be compared directly to those from the other
sources in our sample. The EW and FWHM obtained from more
complex models will always be less than the values obtained from
the empirical modeling so the latter provide useful upper bounds.
He-like and H-like Fe absorption features have been 
reported in the HEG data by Young \etal (2005).

{\it IRAS~13349$+$2438}. This source was observed twice with the
\chandra \hetg but no results for the \fekalfa line have been previously
published. A significant detection of
the narrow \fekalfa line was obtained only from the second 
observation (see \tablesigfixedp).

{\it IC~4329A} McKernan \& Yaqoob (2004) reported the detection of 
complex Fe K line emission from 
the \chandra \hetg observation of this source. 
One peak is centered at $\sim 6.3$~keV with a FWHM 
$20830^{+10110}_{-7375} \rm \ km \ s^{-1}$ and an EW of $110^{+46}_{-40}$~eV.
The other peak is at $\sim 6.9$~keV with a FWHM $\sim 4000 \rm \ km \ s^{-1}$ 
and an EW of $\sim 40$~eV (probably due to \feklyap). 
In the present analysis we are concerned only with the low-ionization 
\fekalfa line. Our re-analysis with the line width fee
is consistent with the results
of McKernan \& Yaqoob (2004) but we note that our 
fits in which the \fekalfa line
width was fixed at well below the HEG 
resolution yielded a line centroid energy
of $6.399^{+0.006}_{-0.005}$~keV. Therefore the \fekalfa 
line parameters from the latter fit are more reliable values for
the true narrow core of the \fekalfa line.
 
{\it Mrk~279}. Results of the new analysis for this source
are consistent with those reported in YP04.

{\it NGC 5506}.
Results from the \chandra \hetg observation
of this source have been presented by Bianchi \etal (2003), 
who obtained FWHM$< 4000 \rm \ km \ s^{-1}$ for
the narrow \fekalfa line at $\sim 6.4$~keV at 99\% confidence. 
We obtained a tighter limit on the FWHM (\tablesigfreep).
Bianchi \etal (2003) did not provide constraints on the line centroid 
energy or EW.

{\it NGC 5548}. Results for both of the \chandra \hetg observations
have already been reported in Yaqoob \etal (2001)
and YP04, and the new analysis is consistent
with the previous results.

{\it Mrk~290}. There are four \chandra \hetg observations for this source
and no results on the \fekalfa line from the HEG data have
previously been published.
None of the individual observations yielded a
detection of the 
narrow \fekalfa line greater than 99\% confidence (for two free 
parameters). However, the line was 
detected with $>3\sigma$ confidence in the time-averaged spectra. 
The line intensity against centroid energy confidence contours shown 
in \figspectot were obtained with the line width fixed at 1~eV
since a closed 99\% confidence contour could not be obtained when the 
line width was a free parameter. 

{\it PDS~456}. Results from the \chandra \hetg observation of this
source 
pertaining to the narrow \fekalfa line have never been previously published.
The signal-to-noise ratio is poor and we could only obtain upper limits
on the EW after fixing the line energy at 6.4~keV.

{\it E1821+643}. Results from the \chandra \hetg observation 
of this source have been presented by Fang et al. (2002) and Yaqoob \& 
Serlemitsos (2005). The latter work
reported \fekalfa line parameters 
\fekecenp$=6.43^{+0.06}_{-0.05}$~keV, 
${\rm EW}=144^{+67}_{-57}$~eV, 
and FWHM$=10980^{+3300}_{-7690} \rm \ km \ s^{-1}$.
However, as described
in Yaqoob \& Serlemitsos (2005),
these parameters are quite model-dependent because an
absorption line was reported at $\sim 6.2$~keV in the rest frame,
and there may also be an underlying broad emission line.
Our fits with the line width fixed at 1~eV likely give the
most representative values of the centroid energy and EW of the
narrow core of the \fekalfa line.

{\it 3C~382}. Gliozzi et al. (2007) have presented the results 
from the two \chandra \hetg observations of this source. 
The \fekalfa line was detected with less than 90\% and 
less than 99\% confidence in first and 
second observations, respectively. From the second observation
Gliozzi \etal (2007) obtained 
\fekecenp$=6.43^{+0.05}_{-0.07}$~keV,
${\rm EW}=55^{+47}_{-20}$~eV, and FWHM$<9560 \rm \ km \ s^{-1}$.
Our results are generally
consistent with those of Gliozzi \etal (2007),
but we note that the latter work also reported results
from the $-1$ and $+1$ orders of the HEG separately, giving
a larger dispersion in the parameter ranges.
 
{\it IRAS~18325$-$5926} and {\it 4C~74.26}.
No results from the \chandra \hetg observations
(pertaining to the \fekalfa line or otherwise) for either of
these sources have
been previously published. 
In IRAS~18325$-$5926 our analysis revealed no significant
detection of the narrow \fekalfa line in either of the two observations
or from the summed spectrum. The detection of the line in 4C~74.26
was marginal even for the spectrum summed over two observations.
Only upper limits on the EW could be derived for IRAS~18325$-$5926
(with the \fekalfa line energy fixed at 6.4~keV).

{\it Mrk~509}. Results from the \chandra \hetg observation of this 
source have been reported in YP04 and the new analysis gives consistent
results.

{\it NGC~7213}. 
Results from the two \chandra \hetg observations of this
source have been reported by Bianchi et al. (2008).
The narrow \fekalfa line parameters obtained were
\fekecen~$=6.397^{+0.006}_{-0.011}$~keV, 
${\rm EW}=120^{+40}_{-30}$~eV, and FWHM$=2400^{+1100}_{-600}
\rm \ km \ s^{-1}$. Our results are consistent with those
of Bianchi \etal (2008), who also reported the detection of
\fexxv and \feklya emission 
lines in the HEG data.

{\it NGC~7314}. Complex Fe~K line emission from multiple ionization 
states was observed by the \chandra  \hetg, and the results of a 
detailed analysis were published by Yaqoob et al. (2003).
The \fekalfa line at $\sim 6.4$~keV is unresolved 
with FWHM$<3520 \rm \ km \ s^{-1}$ and 
and ${\rm EW}=81\pm34$~eV. 
The results presented in the present paper
(\tablesigfixedp) were obtained from fits with the line width
fixed at 1~eV.
Emission lines from 
\fexxv and \feklya have been noted and discussed in detail by
Yaqoob \etal (2003).

{\it Ark~564}. Results pertaining to the narrow \fekalfa line
from the \chandra \hetg observation 
have been presented by Matsumoto, Leighly, \& Marshall (2004) and YP04. The
signal-to-noise ratio of the data is poor and the EW of the line
could only be measured with the line energy fixed at  6.4~keV
and the line width fixed at 1~eV, 
and the results are consistent with those of YP04 (\tablesigfixedp).

{\it MR~2251$-$178}. Gibson et al (2005) reported results from
a \chandra  observation, giving  ${\rm EW}=25\pm13$~eV
and { FWHM$<1530 \rm \ km \ s^{-1}$} for an \fekalfa line 
with a centroid energy fixed at 6.4~keV. 
In our uniform analysis, we found that the line was detected 
at less than 99\% confidence. 
In this case we were not able to obtain constraints on the line width.
Gibson \etal (2005) also reported
the detection of a resolved \feklya absorption line
with ${\rm EW} \sim 28$~eV and a velocity shift of 
$\sim-12700 \rm \ km \ s^{-1}$,
indicating a high-velocity outflow. 

{\it NGC~7469}. Results from the two \chandra \hetg observations have been
reported by Scott et al. (2005). 
A strong \fekalfa line was detected with a centroid energy of $6.39\pm0.01$ keV,
and a line width $6310\pm1580$ km s$^{-1}$. No EW was given but
the line flux was 3.9$\pm0.7\times10^{-5} \rm \ photons \ cm^{-2} \ s^{-1}$.
Our results are consistent (within the statistical
errors) with those of Scott \etal (2005).

\newpage

\section*{Figure Captions}

\par\noindent
{\bf Figure 1} \\
{\it Left~Panels}: \chandra HEG spectra in the Fe K band 
for sources in which the \fekalfa emission line
was detected in only one observation, and which were not included
in the sample of Yaqoob \& Padmanabhan (2004).
The data are binned at $0.01\AA$, comparable to  
to the HEG spectral resolution, which is $0.012\AA$
FWHM. The data are combined from the $-1$ and $+1$ orders
of the grating. The spectra have been corrected for instrumental effective area
and cosmological redshift.
Note that these are {\it not} unfolded spectra and are
therefore independent of any model that is fitted. 
{ Although the spectral fitting was performed using XSPEC, the spectral
plots were {\it not} made using XSPEC.
The statistical errors shown correspond
to the $1\sigma$ 
Poisson errors, which we calculated} using 
equations (7) and (14) in Geherls (1986) that approximate the upper and
lower errors respectively.
The solid line corresponds to a continuum model fitted over the
2--7~keV range (extrapolated to 7.5~keV), 
as described in the text (\S\ref{hegspec}).
The vertical dotted lines represent (from left to right), the rest energies
of the following: Fe~{\sc i}~$K\alpha$, \fexxv forbidden,
two intercombination lines of \fexxvp, \fexxv resonance, \feklyap,
Fe~{\sc i}~$K\beta$, and the Fe~K edge.
{\it Right~Panels}: Joint 99\% 
confidence contours of the \fekalfa emission-line
core intensity versus line centroid energy obtained from 
Gaussian fits to the line
with the line width free as described in the text (solid lines).
For Mrk~590, NGC~985, PG~0844$+$346(1) and IRAS 13349+2438(2), the 
99\% confidence contours (solid lines) of the were poorly
constrained due to the 
intrinsic line width parameter becoming very large. Therefore, 
we overlaid the 99\% confidence contours obtained 
with the line width 
fixed at 1~eV for these cases (dotted contours).
For the remaining sources (ESO~198$-$G24, MCG~$-$6-30-15(2), NGC~5506,
and E1821$+$643), the dotted contours correspond to 68\%, and 90\% confidence.
 
\par\noindent
{\bf Figure 2} \\
{\it Left~Panels}: The time-averaged \chandra HEG spectra in the Fe K band 
for eight AGN in which the \fekalfa emission line was
detected in more than one observation for cases
that were not already reported in Yaqoob \& Padmanabhan (2004).
The data are binned at $0.01\AA$
except for NGC 4151, which is binned at $0.005\AA$. 
The energies of the vertical dotted lines are described in the
caption to \figindspecp.
{\it Right~Panel} Joint 99\% confidence contours of the \fekalfa 
emission-line
core intensity versus line center energy for time-averaged 
and individual spectra. 
Individual observations are shown in different linestyles while the 
time-averaged contours are shown with a solid line. 
The contour shown for Mrk~290 is from the time-averaged spectrum
only, since none of the individual observations 
had sufficient signal-to-noise ratio to obtain well-constrained contours.
For MCG~$-$6-30-15, we show the contour from the time-averaged spectrum
only, since only one of the four observations 
not reported in Yaqoob \& Padmanabhan (2004)
has a significant detection of narrow \fekalfa line, 
and that contour has already been shown in \figindspecp. 


\par\noindent
{\bf Figure 3} \\
{ Distributions of \fekalfa line core centroid
energies constructed in four different ways.
(a) and (b) were made using the results from individual
observations, whereas (c) and (d) were made from
measurements that used spectra averaged from multiple observations
of a given source
where relevant (see \S\ref{properties} for exceptions).
In (a) and (c)
the line intrinsic width was fixed at 1~eV (results from \tablesigfixedp).
In (b) and (d)  
the line centroid energies that could be measured
with the line width free were utilized (i.e. those from
\tablesigfreep), keeping the line measurements from \tablesigfixed
for the remainder.
For the individual observations this results in 51 out of
68 values being obtained with the line width free --see text for details.
}
The dashed and dotted lines in each case
correspond to the distribution of
68\% confidence lower and upper limits on the line 
centroid energy, respectively. 

\par\noindent
{\bf Figure 4} \\
{ Distributions of \fekalfa line EW
constructed in four different ways.
(a) and (b) were made using the results from individual
observations, whereas (c) and (d) were made from
measurements that used spectra averaged from multiple observations
of a given source
where relevant (see \S\ref{properties} for exceptions).
In (a) and (c)
the line intrinsic width was fixed at 1~eV (results from \tablesigfixedp).
In (b) and (d) 
the line centroid energies that could be measured  
with the line width free were utilized (i.e. those from
\tablesigfreep), keeping the line measurements from \tablesigfixed
for the remainder.
For the individual observations this resulted in 70 out of
82 values being obtained with the line width free --see text for details.}
The dashed and dotted lines in each case
correspond to the distribution of
68\% confidence lower and upper limits on the line 
EW, respectively.
The shaded histograms in both panels 
mark the 68\% upper limits on the EW for 12 
observations in which the EW could not be measured.
Note that the largest EW of $\sim600$~eV 
(for PG~0844$+$349) is not shown in (b) and (d) because
it is not a true measure of the narrow-line EW at $\sim 6.4$~keV
(see text).

\par\noindent
{\bf Figure 5} \\
The \fekalfa emission-line FWHM versus the H$\beta$ 
FWHM for which the \fekalfa line 
width could be constrained (see text and \tablesigfreep). 
For MCG~$-$5-23-16, we used the FWHM of infra-red broad Br$\alpha$ line as
a surrogate for H$\beta$ FWHM.
The dashed line corresponds to the two line widths being 
equal. Open circles correspond to the 12 cases
shown in \figifevsfwhmp,
for which the best \fekalfa line FWHM constraints were obtained
(see text).
The statistical errors on the \fekalfa line FWHM shown 
correspond to 68\% confidence 
for three free parameters. 

\par\noindent
{\bf Figure 6} \\
Joint 68\%, 90\%, and 99\%, confidence contours of the \fekalfa emission-line
core EW versus the ratio of the \fekalfa FWHM to the H$\beta$ FWHM for 
12 AGN
that provided the best measurements of \fekalfa line FWHM (see text).
For MCG~$-$5-23-16, we used the FWHM of infra-red broad Br$\alpha$ line as
a surrogate for H$\beta$ FWHM.
The vertical dotted lines correspond to a FWHM ratio
of the pairs of emission lines equal to unity.

\par\noindent
{\bf Figure 7} \\
{ (a) The \fekalfa core emission-line EW versus 
the 2--10 keV luminosity. (b) As (a)
for EW versus
$(L_{\rm 2-10 \ keV}/L_{\rm Edd})$, a proxy for the accretion rate.
Both (a) and (b) were constructed from measurements made from individual
observations. 
(c) As (a) but showing EW versus
$L_{\rm 2-10 \ keV}$  for measurements made from spectra combining
multiple observations for a given source, where relevant.
(d) As (c) but showing EW versus $(L_{\rm 2-10 \ keV}/L_{\rm Edd})$.
In (c) and (d) the average spectrum was not used for all sources,
for (b) and (d) reliable black-hole mass estimates were not available
for all sources-
see \S\ref{properties} and \S\ref{baldwineffect} for details. 
All of the measurements shown in (a)--(d)
utilize results from the spectral fitting in
which the \fekalfa line intrinsic width was fixed at 1~eV. 
The statistical errors on the \fekalfa line EW correspond to 68\% confidence.
The dotted lines show the correlations obtained
by linear fits to  
$\log{\rm EW}$ versus $\log{L_{\rm 2-10 \ keV}}$ (a) and (c), and 
$\log{\rm EW}$ versus
$\log{(L_{\rm 2-10 \ keV}/L_{\rm Edd})}$ (b) and (d).}
Note that observations with only upper limits on the EW were
not included in the fits.

\begin{figure*}[h]
\vspace{10pt}
\centerline{\psfig{file=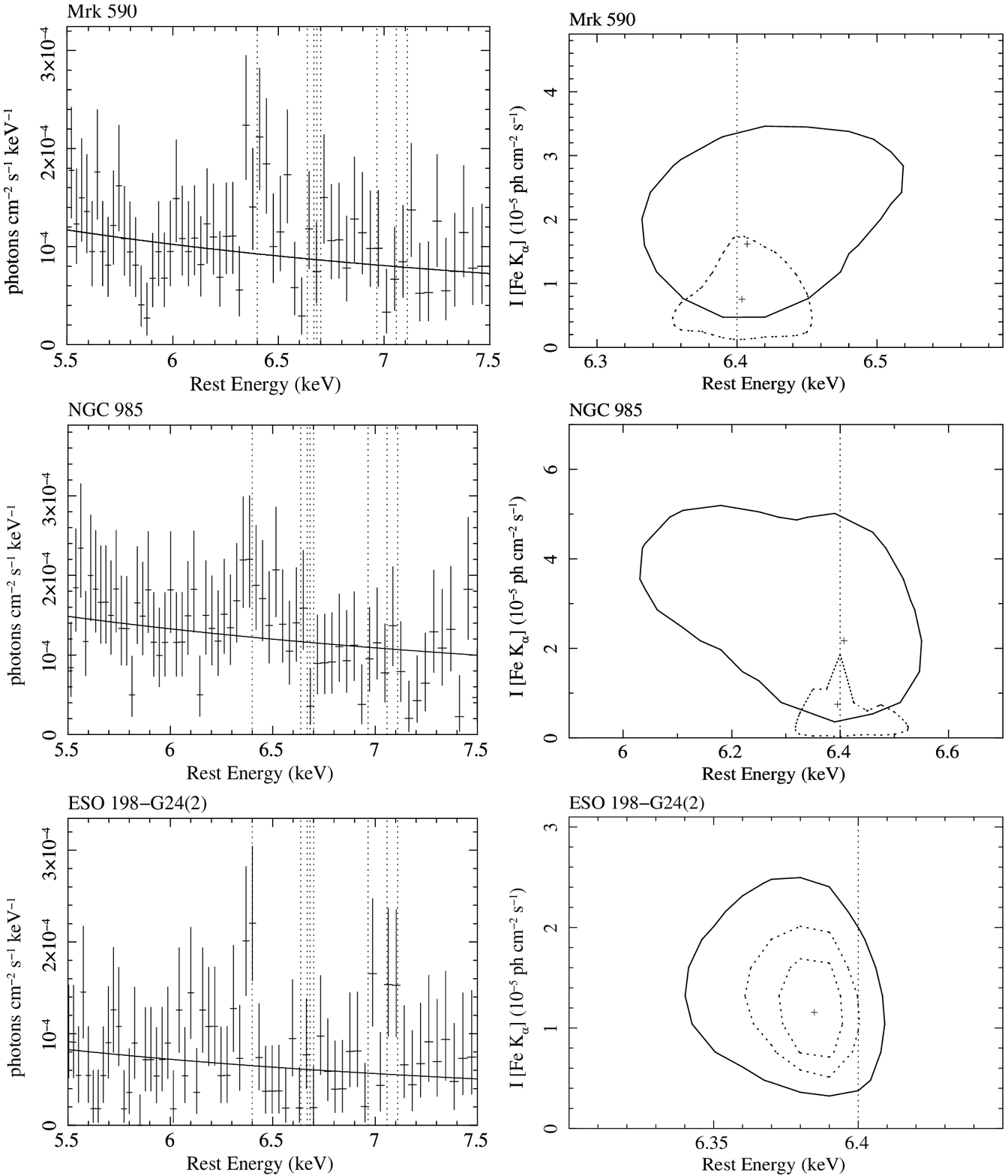,width=6.5in,height=8.in,angle=0}}
\caption{\footnotesize
}
\end{figure*}

\setcounter{figure}{0}
\begin{figure*}[tbh]
\vspace{10pt}
\centerline{\psfig{file=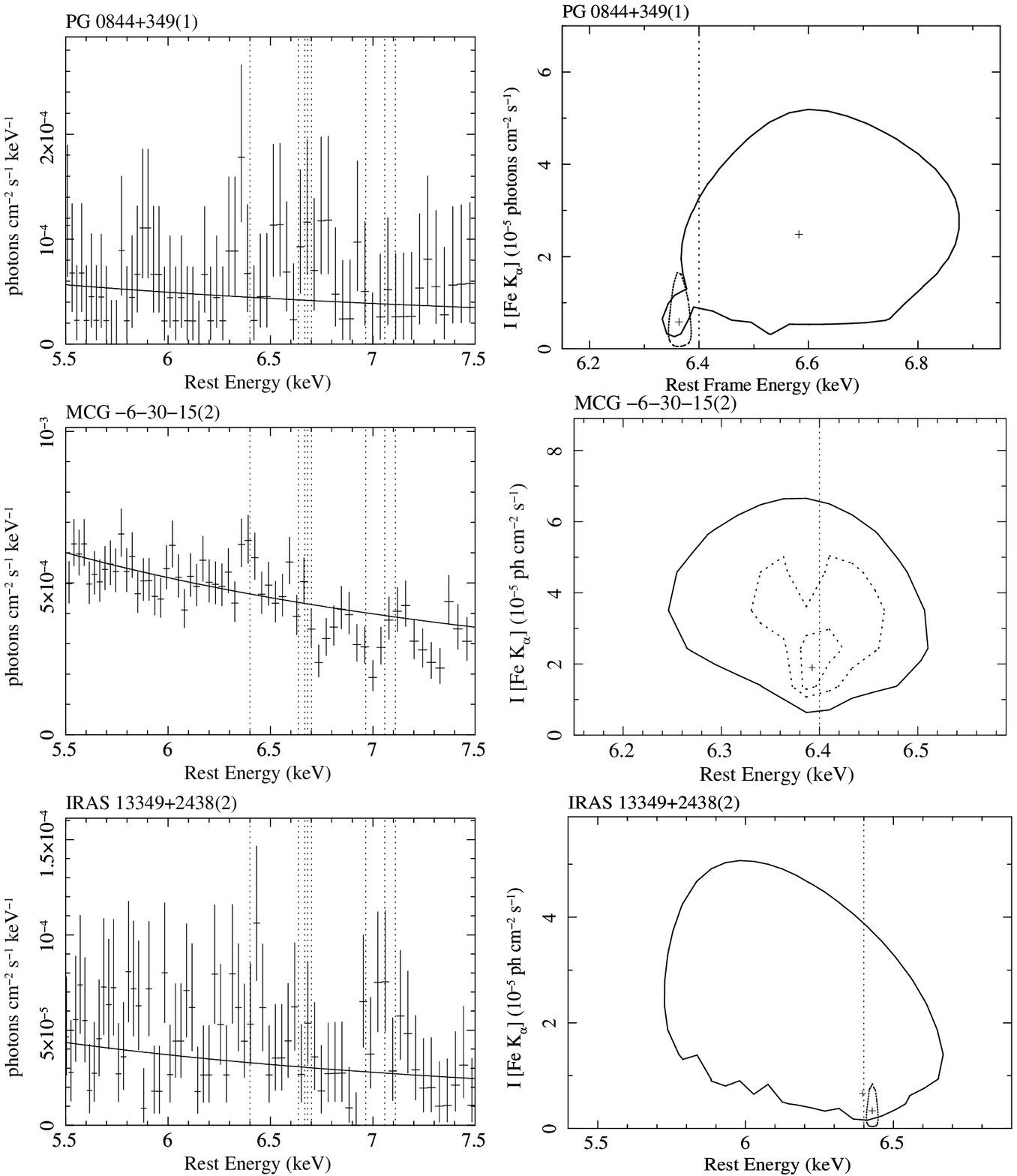,width=6.5in,height=8in,angle=0}}
\caption{ -- {\it continued}}
\end{figure*}

\setcounter{figure}{0}
\begin{figure*}[tbh]
\vspace{10pt}
\centerline{\psfig{file=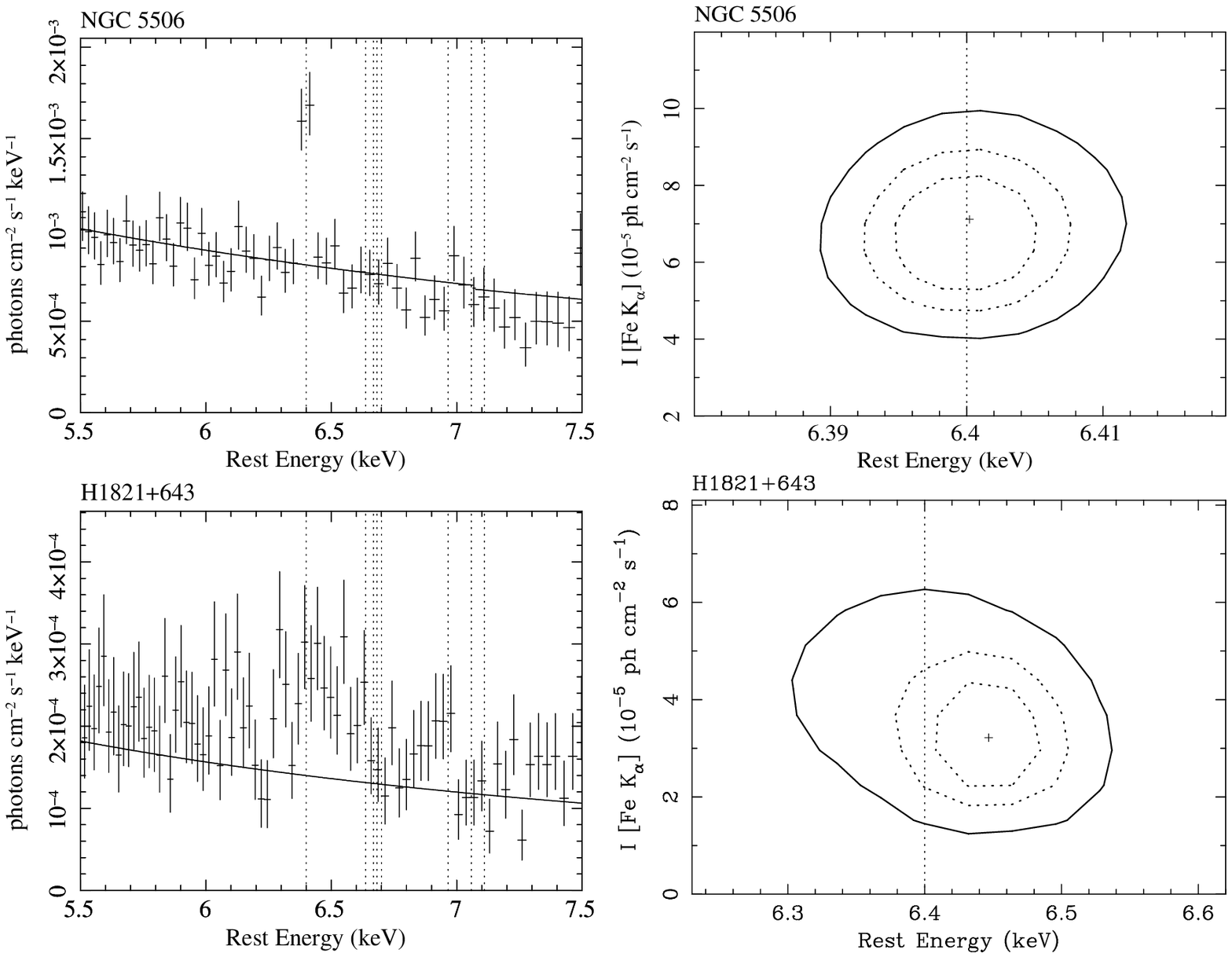,width=6.5in,height=5.5in,angle=0}}
\caption{ -- {\it continued}}
\end{figure*}

\newpage

\begin{figure*}[h]
\vspace{10pt}
\centerline{\psfig{file=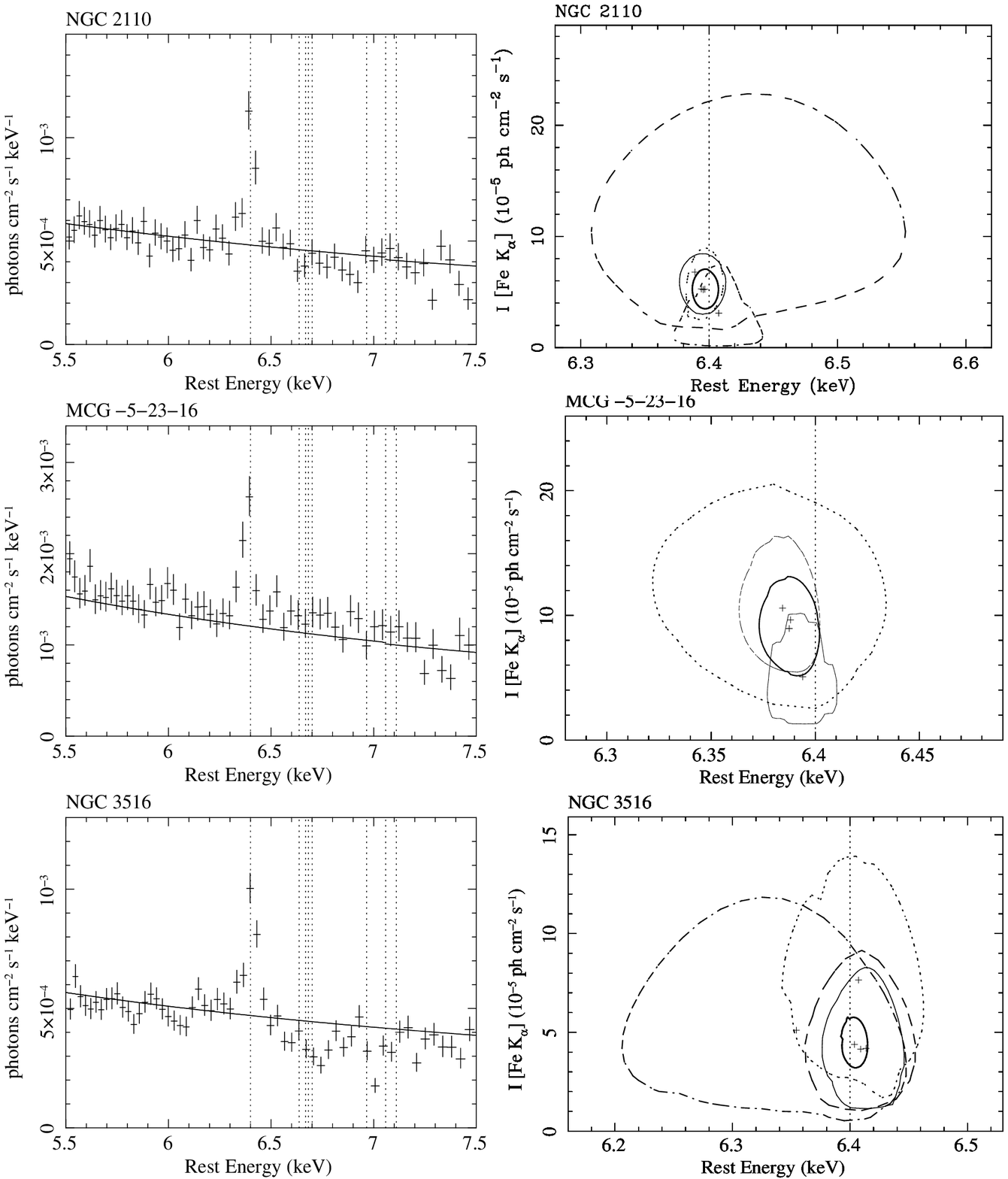,width=6.5in,height=8.in,angle=0}}
\caption{\footnotesize
}
\end{figure*}

\setcounter{figure}{1}
\begin{figure*}[tbh]
\vspace{10pt}
\centerline{\psfig{file=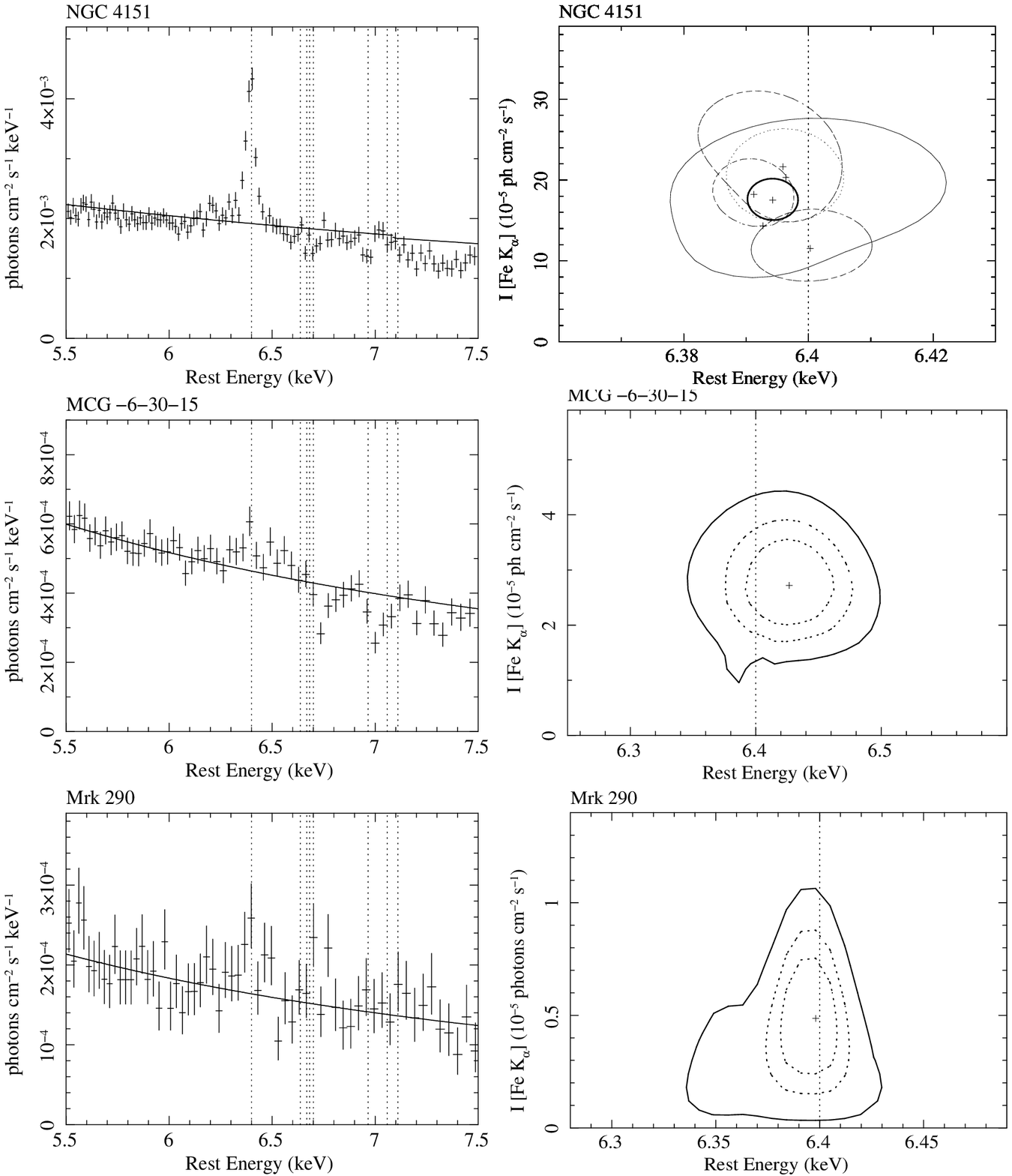,width=6.5in,height=8.in,angle=0}}
\caption{ -- {\it continued}}
\end{figure*}

\setcounter{figure}{1}
\begin{figure*}[tbh]
\vspace{10pt}
\centerline{\psfig{file=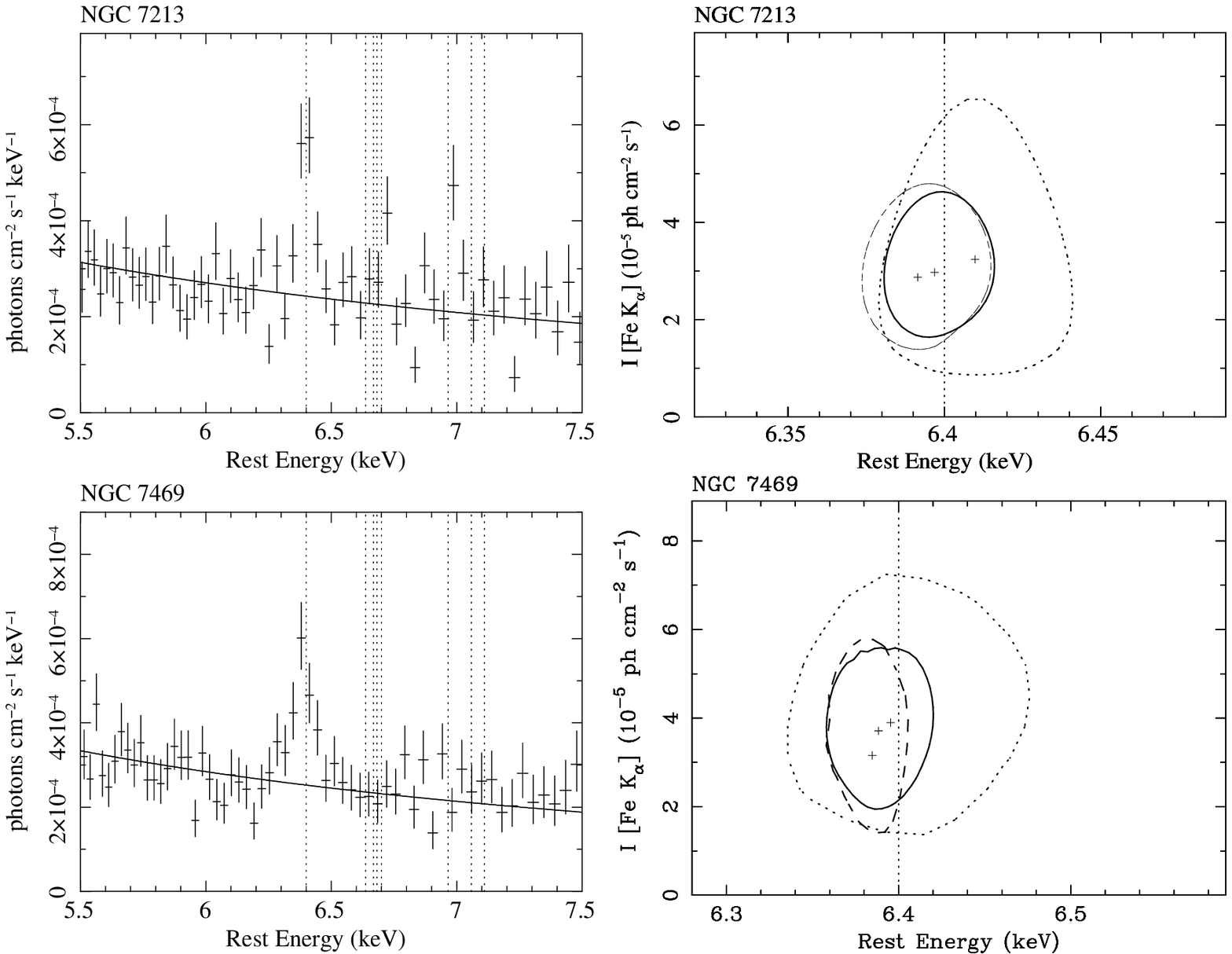,width=6.5in,height=5.5in,angle=0}}
\caption{ -- {\it continued}}
\end{figure*}

\begin{figure}
\epsscale{2.0}
\centerline{\psfig{file=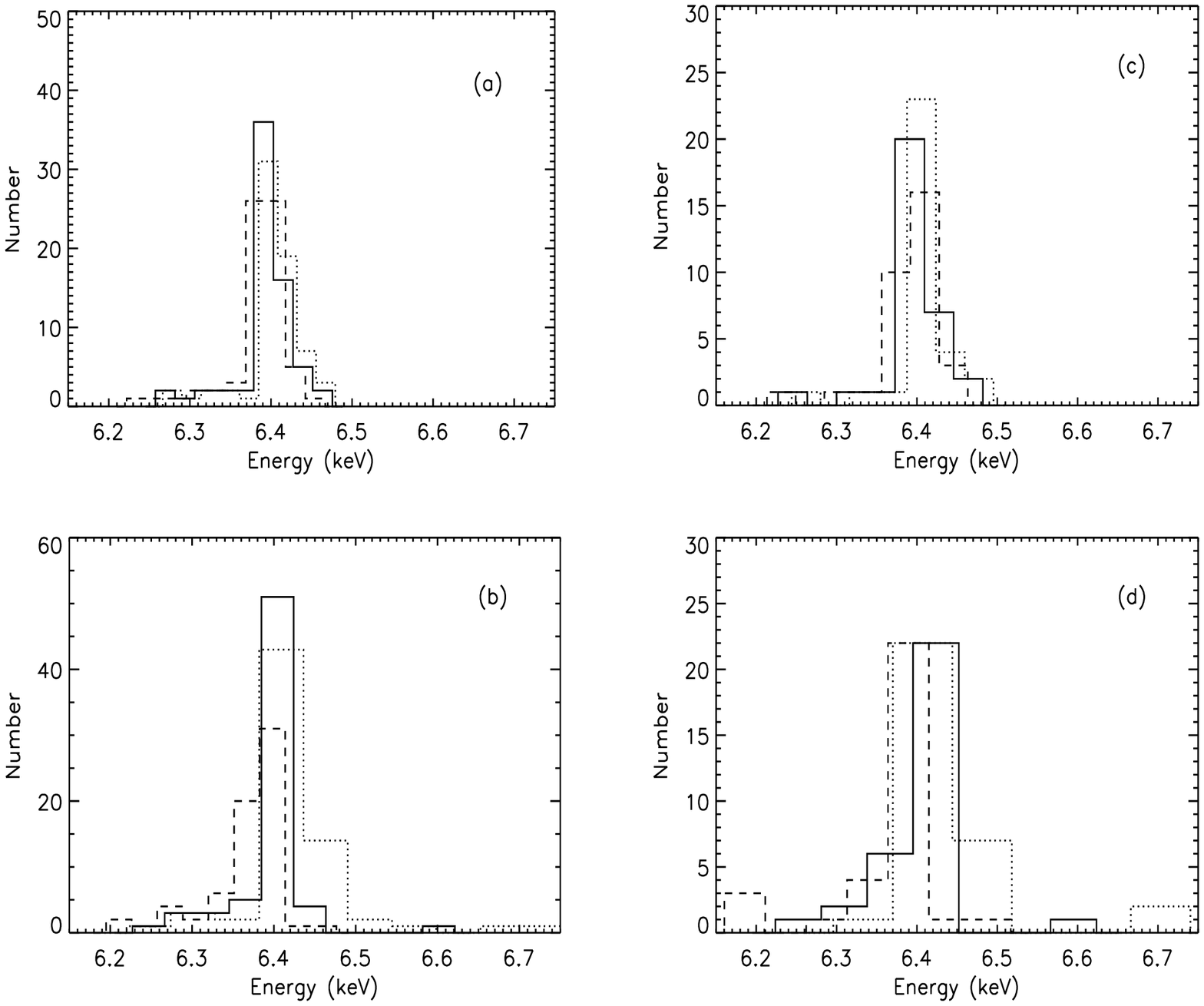,width=8.0in,height=5.5in,angle=0}}
\caption{ }
\end{figure}

\begin{figure}
\epsscale{2.0}
\centerline{\psfig{file=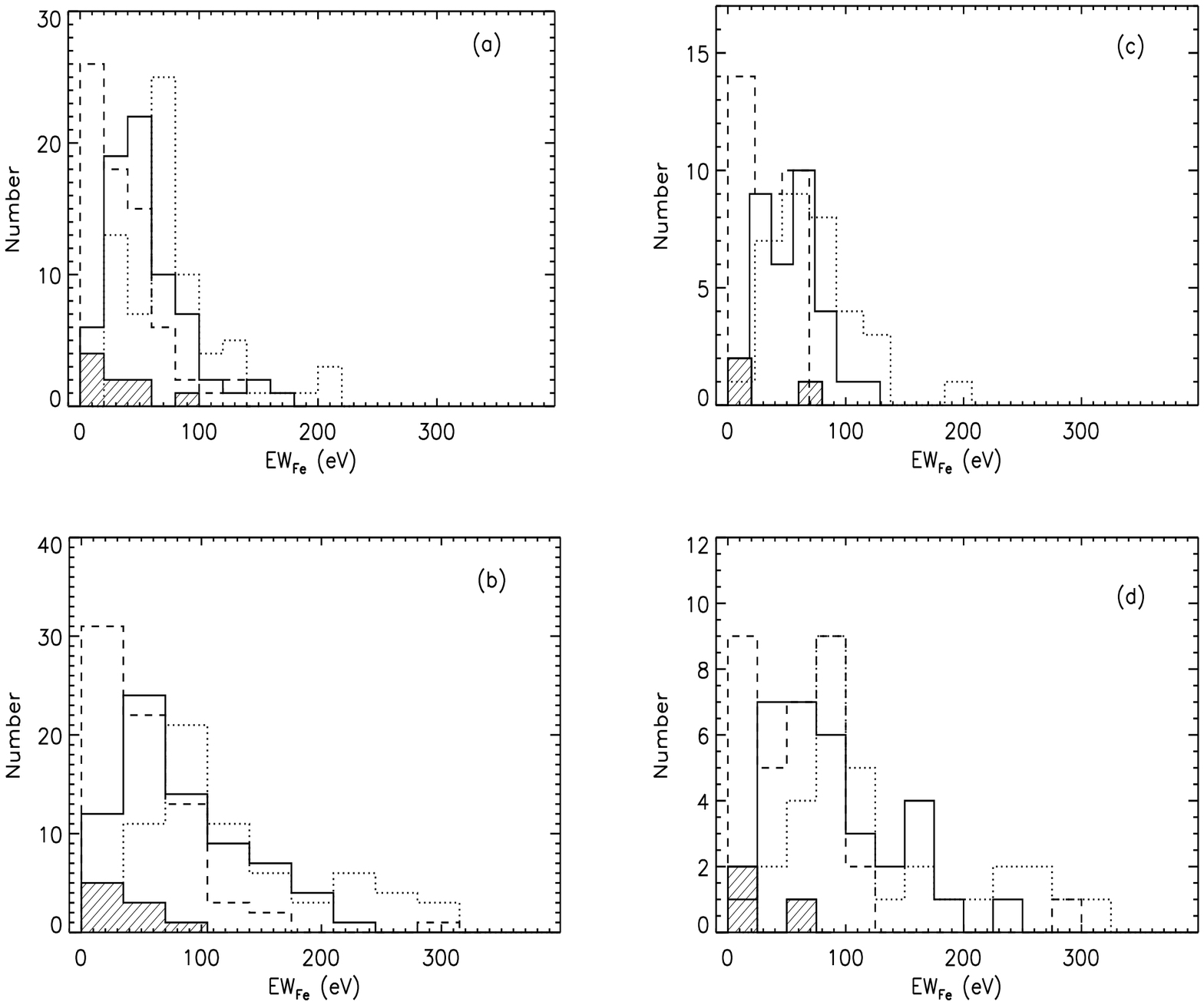,width=8.0in,height=5.5in,angle=0}}
\caption{ }
\end{figure}

\begin{figure}
\epsscale{1.0}
\plotone{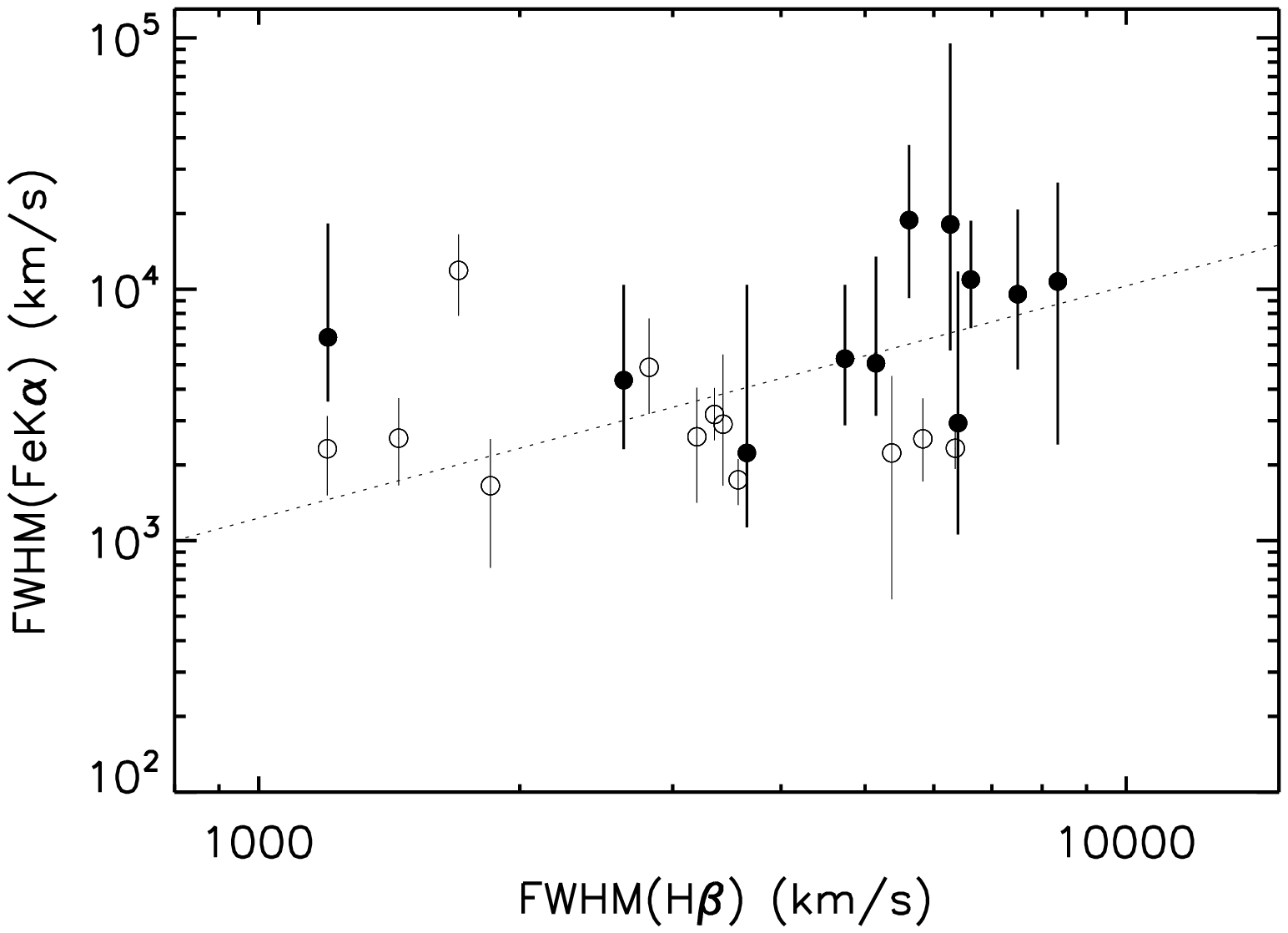}
\caption{ }
\end{figure}

\begin{figure*}[h]
\vspace{10pt}
\centerline{\psfig{file=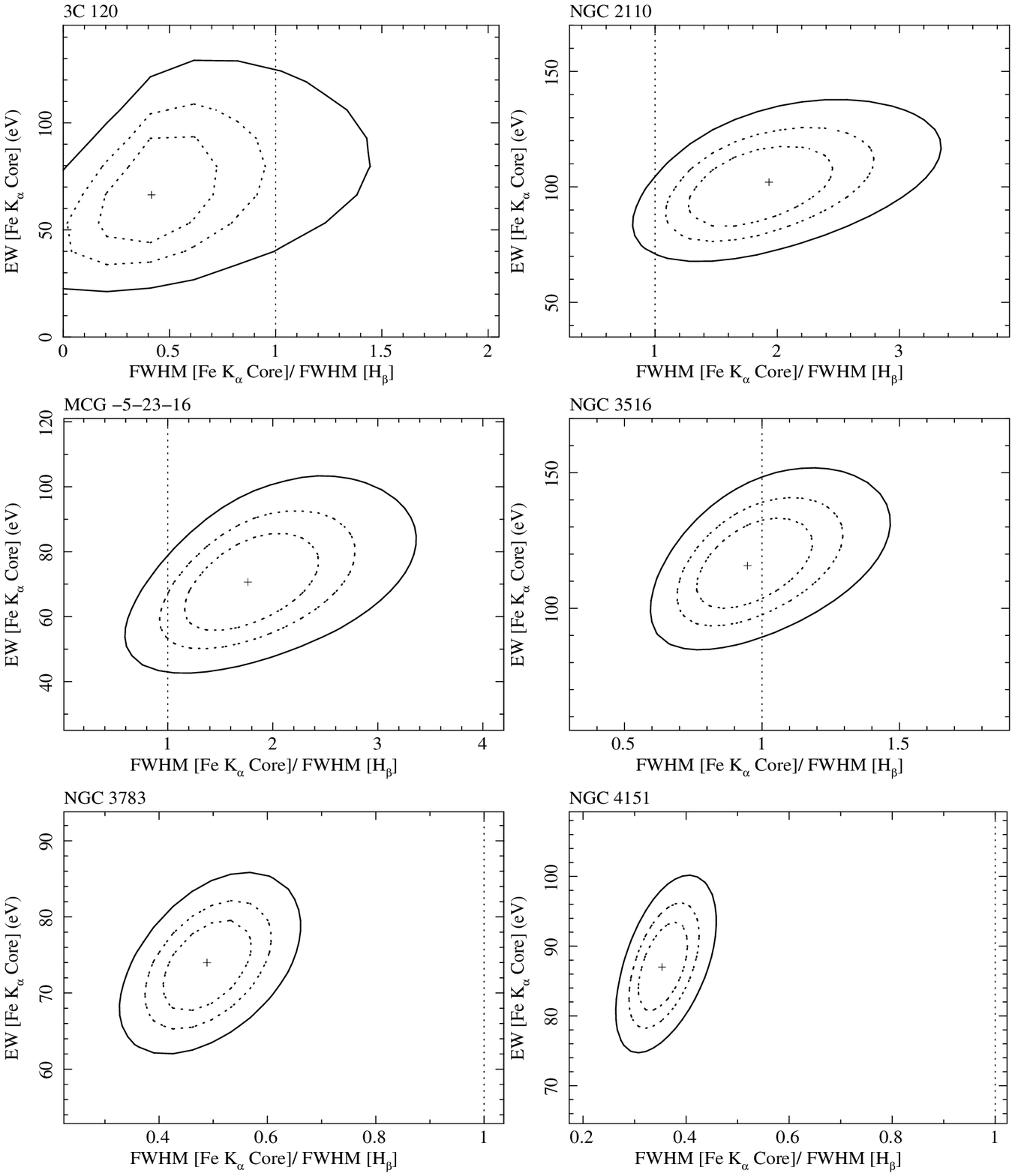,width=6.5in,height=8.in,angle=0}}
\caption{\footnotesize
}
\end{figure*}

\setcounter{figure}{5}
\begin{figure*}[tbh]
\vspace{10pt}
\centerline{\psfig{file=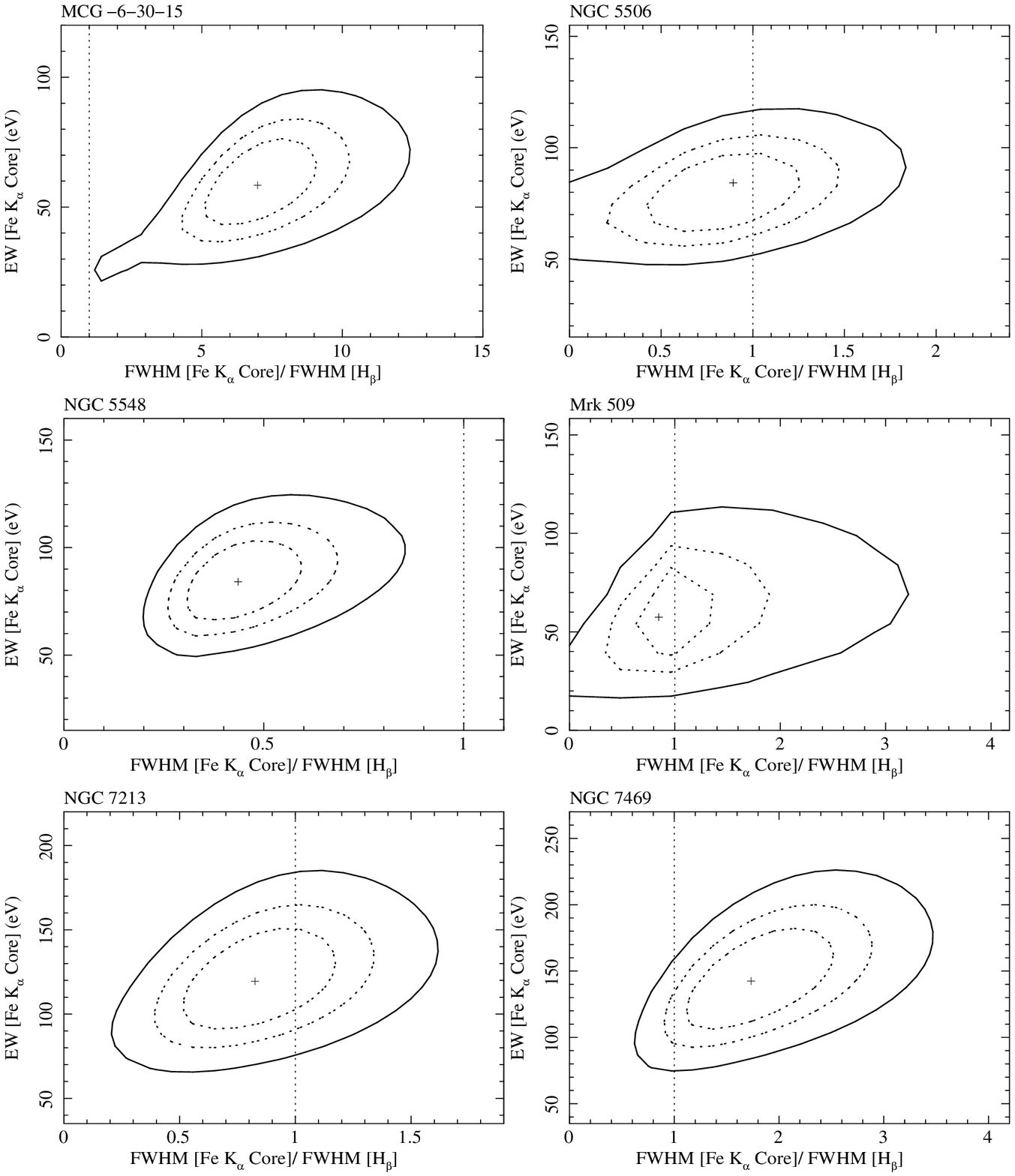,width=6.5in,height=8.in,angle=0}}
\caption{ -- {\it continued}}
\end{figure*}

\begin{figure}
\epsscale{2.0}
\centerline{\psfig{file=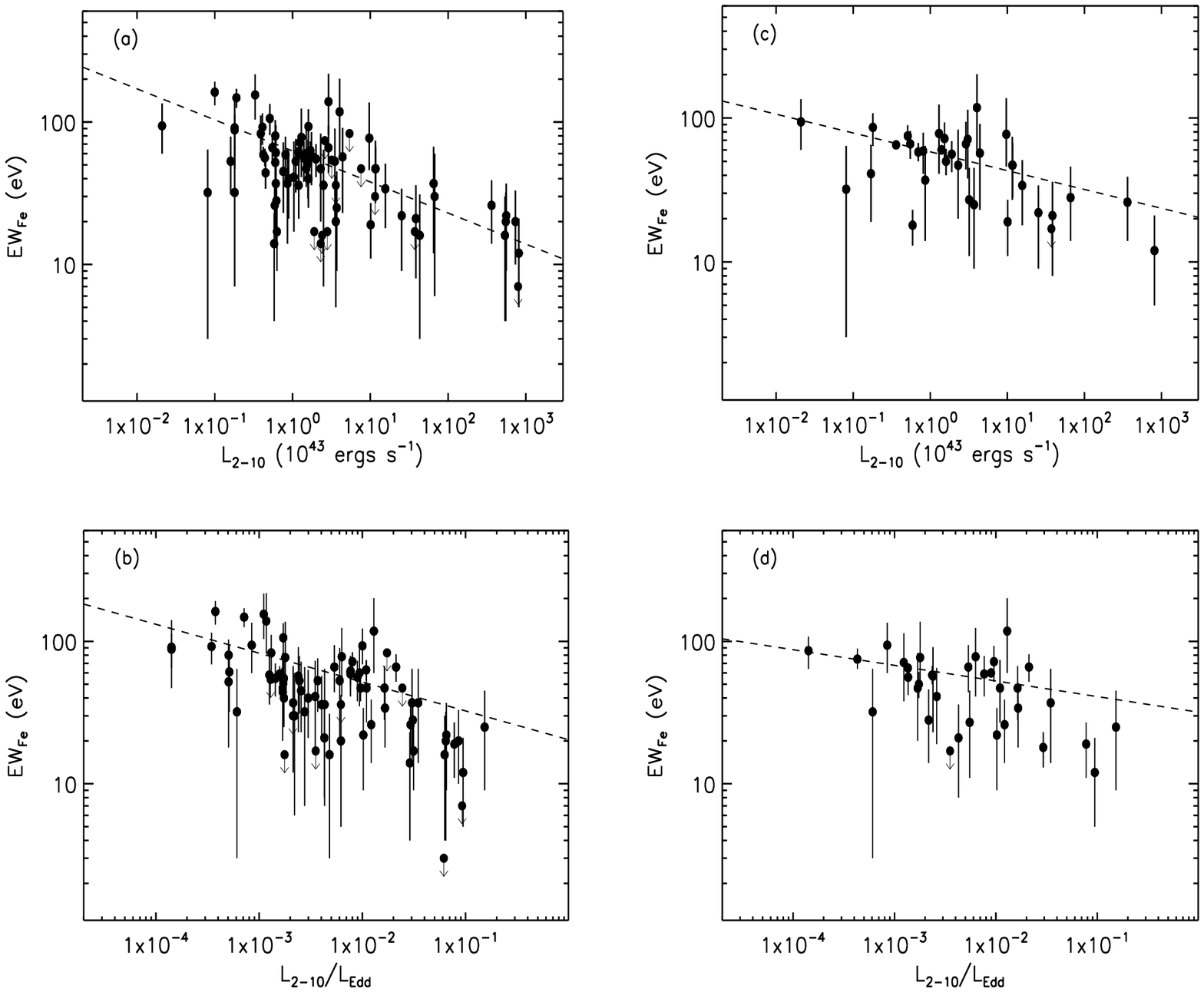,width=8.0in,height=5.5in,angle=0}}
\caption{ 
.\label{fig2}}
\end{figure}

\newpage
\begin{deluxetable}{lcccccccc}
\tabletypesize{\scriptsize}
\tablecaption{\scshape Parameters of the Core Fe K Line Emission ($\sigma$=1 eV) from $\it Chandra$ (HEG) Data}
\tablewidth{0pt}
\tablehead{\colhead{Source} & \colhead{$z$} & \colhead{Seq. Num} &\colhead{$E$} & \colhead{$I$} & \colhead{$EW$}& \colhead{$F$} & \colhead{$L$} & \colhead{$\Delta C$}\\
\hspace*{15.mm} & & /ObsID/exp &(keV) &  &(eV) & (2$-$10 keV) &(2$-$10 keV)   & \\
\hspace*{15.mm} (1)&(2) & (3) & (4) & (5) & (6) & (7) & (8)  & (9)  }
\startdata
 Fairall 9 &    0.0470160&$700278$ &$6.458_{-0.016}^{+0.008}$&$1.2_{-0.5}^{+0.7}$&$47_{-20}^{+27}$&2.2&11.6&14.4\\
            &&$/2088/79.9$&$(6.434-6.468)$&$(0.5-2.2)$&$(20-86)$&&&\\
NGC 526a(1) &    0.0190970&$700840$&$6.400^f$&$0_{-0}^{+0.6}$&$0_{-0}^{+16}$&3.0&2.4&0\\
            &&$/4376/29.1$&$\dots$&$(0-1.0)$&$(0-27)$&&&\\
NGC 526a(2) &    0.0190970&$700840$&$6.400_{-0.006}^{+0.010}$&$1.6_{-0.9}^{+1.3}$&$47_{-27}^{+36}$&3.0&2.3&8.0\\
            &&$/4437/29.4$&$(6.389-6.413)$&$(0.3-3.5)$&$(9.0-100)$&&&\\
NGC 526a(total) &    0.0190970&$\dots$&$6.394_{-0.006}^{+0.012}$&$1.0_{-0.7}^{+0.7}$&$28_{-20}^{+20}$&2.9&2.4&5.0\\
            &&$\dots/57.8$&$(6.380-6.414)$&$(0-2.1)$&$(0-59)$&&&\\
 Mrk 590 &    0.0263850&$701005$&$6.403_{-0.009}^{+0.016}$&$0.8_{-0.4}^{+0.4}$&$78_{-37}^{+46}$&0.85&1.3&14.9\\
            &&$/4924/96.8$&$(6.386-6.435)$&$(0.3-1.5)$&$(31-155)$&&&\\
NGC 985 &    0.0431430&$700449$&$6.395_{-0.009}^{+0.015}$&$0.7_{-0.4}^{+0.5}$&$57_{-34}^{+34}$&1.1&4.4&10.1\\
            &&$/3010/77.7$&$(6.379-6.412)$&$(0.2-1.5)$&$(15-113)$&&&\\
ESO 198$-$G24(1) &    0.0455000&$700900$&$6.400^f$&$0.2_{-0.2}^{+0.2}$&$26_{-26}^{+26}$&0.67&3.2&1.6\\
            &&$/4817/80.3$&$\dots$&$(0-0.6)$&$(0-74)$&&&\\
ESO 198$-$G24(2) &    0.0455000&$700900$&$6.386_{-0.008}^{+0.008}$&$1.0_{-0.4}^{+0.6}$&$139_{-57}^{+79}$&0.61&2.9&23.2\\
            &&$/5315/71.5$&$(6.377-6.401)$&$(0.4-1.8)$&$(55-246)$&&&\\
ESO 198$-$G24(total) &    0.0455000&$\dots$&$6.394_{-0.009}^{+0.008}$&$0.6_{-0.3}^{+0.3}$&$71_{-33}^{+43}$&0.64&3.1&17.3\\
    &&$\dots/151.5$&$(6.377-6.409)$&$(0.2-1.0)$&$(25-126)$&&&\\
3C 120 &   0.00330100&$700454$&$6.412_{-0.009}^{+0.009}$&$2.5_{-1.0}^{+1.1}$&$47_{-19}^{+20}$&4.7&11.7&20.1\\
            &&$/3015/58.2$&$(6.396-6.428)$&$(1.2-4.2)$&$(22-78)$&&&\\
NGC 2110(1) &   0.00778900&$700582$&$6.416_{-0.010}^{+0.008}$&$3.7_{-1.8}^{+2.0}$&$61_{-29}^{+34}$&4.5&0.61&14.1\\
            &&$/3143/34$&$(6.397-6.430)$&$(1.3-6.8)$&$(22-113)$&&&\\
NGC 2110(2) &   0.00778900&$700582$&$6.407_{-0.016}^{+0.015}$&$3.1_{-2.0}^{+2.5}$&$52_{-34}^{+41}$&4.5&0.60&10.4\\
            &&$/3417/33.2$&$(6.384-6.434)$&$(0.6-6.6)$&$(10-110)$&&&\\
NGC 2110(3) &   0.00778900&$700582$&$6.392_{-0.002}^{+0.007}$&$4.9_{-1.3}^{+1.4}$&$80_{-21}^{+23}$&4.5&0.60&51.5\\
            &&$/3418/76.1$&$(6.384-6.400)$&$(3.1-7.0)$&$(51-114)$&&&\\
NGC 2110(4) &   0.00778900&$700841$&$6.392_{-0.002}^{+0.008}$&$3.9_{-0.9}^{+0.9}$&$92_{-23}^{+23}$&3.1&0.41&70.8\\
            &&$/4377/96.4$&$(6.384-6.400)$&$(2.6-5.3)$&$(62-127)$&&&\\
NGC 2110(total) &   0.00778900&$\dots$&$6.399_{-0.008}^{+0.001}$&$3.9_{-0.6}^{+0.7}$&$75_{-11}^{+14}$&3.8&0.51&134\\
            &&$\dots/200.4$&$(6.391-6.400)$&$(3.0-4.9)$&$(58-95)$&&&\\
PG 0844$+$349(1) &    0.0640000&$701023$&$6.364_{-0.009}^{+0.007}$&$0.6_{-0.3}^{+0.4}$&$118_{-58}^{+83}$&0.42&4.0&12.4\\
            &&$/5599/57.2$&$(6.352-6.375)$&$(0.2-1.3)$&$(40-262)$&&&\\
PG 0844$+$349(2) &    0.0640000&$701023$&$6.400^f$&$0.2_{-0.2}^{+0.3}$&$36_{-36}^{+47}$&0.55&5.4&1.3\\
            &&$/6244/50.2$&$\dots$&$(0-0.7)$&$(0-116)$&&&\\
PG 0844$+$349(3) &    0.0640000&$701023$&$6.400^f$&$0.1_{-0.1}^{+0.3}$&$13_{-13}^{+34}$&0.77&7.6&0.2\\
            &&$/6245/36.2$&$\dots$&$(0-0.6)$&$(0-76)$&&&\\
PG 0844$+$349(total) &    0.0640000&$\dots$&$6.366_{-0.008}^{+0.008}$&$0.3_{-0.1}^{+0.3}$&$52_{-20}^{+43}$&0.55&5.4&10.2\\
            &&$\dots/141.2$&$(6.356-6.381)$&$(0.1-0.7)$&$(16-112)$&&&\\
Mrk 705 &    0.0291500&$700995$&$6.400^f$&$0.4_{-0.4}^{+0.6}$&$26_{-26}^{+48}$&1.3&2.6&0.5\\
            &&$/4914/21.3$&$\dots$&$(0-1.6)$&$(0-113)$&&&\\
MCG $-$5-23-16(1) &   0.00827900&$700311$&$6.394_{-0.007}^{+0.007}$&$7.2_{-1.9}^{+2.1}$&$55_{-14}^{+18}$&10.5&1.6&47\\
            &&$/2121/76.2$&$(6.386-6.402)$&$(4.6-10)$&$(35-81)$&&&\\
MCG $-$5-23-16(2) &   0.00827900&$701171$&$6.394_{-0.008}^{+0.008}$&$5.1_{-2.0}^{+2.5}$&$40_{-15}^{+20}$&10.6&1.6&18.3\\
            &&$/6187/30.1$&$(6.384-6.403)$&$(2.3-8.6)$&$(18-68)$&&&\\
MCG $-$5-23-16(3) &   0.00827900&$701171$&$6.395_{-0.010}^{+0.015}$&$6.4_{-2.8}^{+3.0}$&$51_{-22}^{+24}$&10.5&1.6&16.7\\
            &&$/7240/20.3$&$(6.371-6.411)$&$(2.6-10)$&$(21-87)$&&&\\
MCG $-$5-23-16(total) &   0.00827900&$\dots$&$6.394_{-0.001}^{+0.002}$&$6.4_{-1.3}^{+1.4}$&$50_{-10}^{+11}$&10.5&1.6&80.9\\
            &&$\dots/96.1$&$(6.387-6.402)$&$(4.6-8.4)$&$(36-65)$&&&\\
NGC 3227 &   0.00385900&$700165$&$6.388_{-0.012}^{+0.021}$&$1.0_{-0.9}^{+1.0}$&$32_{-29}^{+32}$&2.5&0.08&3.1\\
            &&$/860/47$&$\dots$&$(0-2.6)$&$(0-83)$&&&\\
NGC 3516(1) &   0.00883600&$700270$&$6.395_{-0.006}^{+0.002}$&$3.8_{-0.9}^{+1.0}$&$106_{-25}^{+28}$&3.0&0.51&69.3\\
            &&$/2080/74.5$&$(6.383-6.398)$&$(2.5-5.2)$&$(70-145)$&&&\\
NGC 3516(2) &   0.00883600&$700270$&$6.406_{-0.008}^{+0.009}$&$3.7_{-1.2}^{+1.5}$&$155_{-51}^{+61}$&1.9&0.33&38.2\\
            &&$/2431/36.2$&$(6.397-6.416)$&$(2.1-5.9)$&$(87-246)$&&&\\
NGC 3516(3) &   0.00883600&$700270$&$6.398_{-0.008}^{+0.002}$&$2.4_{-0.7}^{+0.9}$&$83_{-26}^{+28}$&2.3&0.39&40.6\\
            &&$/2482/89.5$&$(6.389-6.407)$&$(1.4-3.6)$&$(47-121)$&&&\\
NGC 3516(4) &   0.00883600&$701337$&$6.407_{-0.010}^{+0.016}$&$2.5_{-1.2}^{+1.4}$&$40_{-19}^{+23}$&5.2&0.89&12.5\\
            &&$/7281/43.1$&$(6.373-6.430)$&$(0.8-4.6)$&$(13-74)$&&&\\
NGC 3516(5) &   0.00883600&$701337$&$6.398_{-0.008}^{+0.009}$&$2.3_{-1.1}^{+1.4}$&$45_{-22}^{+27}$&4.4&0.76&12.9\\
            &&$/7282/42.1$&$(6.383-6.414)$&$(0.8-4.3)$&$(16-83)$&&&\\
NGC 3516(6) &   0.00883600&$701337$&$6.430_{-0.008}^{+0.008}$&$3.8_{-1.5}^{+1.6}$&$53_{-21}^{+23}$&6.5&1.1&20.6\\
            &&$/8450/39.1$&$(6.415-6.439)$&$(1.7-6.2)$&$(24-87)$&&&\\
NGC 3516(7) &   0.00883600&$701337$&$6.407_{-0.010}^{+0.030}$&$2.7_{-1.1}^{+1.6}$&$36_{-15}^{+21}$&6.7&1.2&14.4\\
            &&$/8451/48.1$&$(6.389-6.439)$&$(1.1-4.9)$&$(15-65)$&&&\\
NGC 3516(8) &   0.00883600&$701337$&$6.431_{-0.033}^{+0.009}$&$2.9_{-1.8}^{+2.2}$&$41_{-24}^{+32}$&6.1&1.0&7.2\\
            &&$/8452/20.2$&$(6.390-6.447)$&$(0.5-6.2)$&$(7.0-88)$&&&\\
NGC 3516(total) &   0.00883600&$\dots$&$6.398_{-0.001}^{+0.001}$&$2.8_{-0.4}^{+0.4}$&$58_{-8}^{+9}$&4.1&0.70& 161.7\\
            &&$\dots/386.5$&$(6.397-6.400)$&$(2.2-3.4)$&$(46-71)$&&&\\
NGC 3783(1) &   0.00973000&$700045$&$6.396_{-0.008}^{+0.007}$&$4.4_{-1.4}^{+1.3}$&$56_{-18}^{+17}$&6.8&1.4&36.6\\
            &&$/373/57.2$&$(6.387-6.404)$&$(2.5-6.4)$&$(32-88)$&&&\\
NGC 3783(2) &   0.00973000&$700280$&$6.403_{-0.008}^{+0.001}$&$4.2_{-0.8}^{+0.8}$&$61_{-12}^{+11}$&5.9&1.2&105.2\\
            &&$/2090/167.8$&$(6.395-6.404)$&$(3.1-5.4)$&$(45-78)$&&&\\
NGC 3783(3) &   0.00973000&$700281$&$6.395_{-0.001}^{+0.001}$&$4.3_{-0.8}^{+0.8}$&$62_{-11}^{+12}$&5.9&1.2&109.2\\
            &&$/2091/171$&$(6.387-6.397)$&$(3.2-5.5)$&$(46-79)$&&&\\
NGC 3783(4) &   0.00973000&$700282$&$6.396_{-0.001}^{+0.001}$&$5.0_{-0.8}^{+0.8}$&$72_{-11}^{+12}$&6.0&1.3&146.6\\
            &&$/2092/167.6$&$(6.394-6.401)$&$(3.9-6.2)$&$(56-90)$&&&\\
NGC 3783(5) &   0.00973000&$700283$&$6.396_{-0.001}^{+0.001}$&$5.8_{-0.9}^{+0.9}$&$63_{-10}^{+10}$&8.2&1.7&154.2\\
            &&$/2093/168.2$&$(6.394-6.397)$&$(4.5-7.1)$&$(49-77)$&&&\\
NGC 3783(6) &   0.00973000&$700284$&$6.396_{-0.001}^{+0.008}$&$4.0_{-0.8}^{+0.8}$&$48_{-9}^{+10}$&7.2&1.5&89.3\\
            &&$/2094/168.3$&$(6.394-6.404)$&$(2.9-5.2)$&$(35-62)$&&&\\
NGC 3783(total) &   0.00973000&$\dots$&$6.396_{-0.001}^{+0.001}$&$4.6_{-0.3}^{+0.4}$&$60_{-4}^{+5}$&6.6&1.4& 635.9\\
            &&$\dots/888.7$&$(6.395-6.396)$&$(4.1-5.1)$&$(53-66)$&&&\\
NGC 4051 &   0.00233600&$700164$&$6.414_{-0.010}^{+0.006}$&$1.6_{-0.5}^{+0.9}$&$94_{-34}^{+41}$&1.7&0.02&26.5\\
            &&$/859/80.8$&$(6.398-6.422)$&$(0.9-2.8)$&$(49-152)$&&&\\
NGC 4151(1) &   0.00331900&$700007$&$6.396_{-0.001}^{+0.001}$&$17.2_{-2.6}^{+2.7}$&$148_{-22}^{+23}$&7.8&0.19&202.3\\
            &&$/335/48$&$(6.395-6.398)$&$(13-21)$&$(117-182)$&&&\\
NGC 4151(2) &   0.00331900&$700491$&$6.396_{-0.001}^{+0.001}$&$13.9_{-1.4}^{+1.6}$&$59_{-6}^{+8}$&17.4&0.42& 306.5\\
            &&$/3052/156.6$&$(6.394-6.396)$&$(11-16)$&$(50-69)$&&&\\
NGC 4151(3) &   0.00331900&$700491$&$6.396_{-0.001}^{+0.001}$&$14.0_{-1.9}^{+2.1}$&$56_{-7}^{+9}$&18.3&0.44&177.1\\
            &&$/3480/92.9$&$(6.394-6.397)$&$(11-17)$&$(46-69)$&&&\\
NGC 4151(4) &   0.00331900&$701493$&$6.396_{-0.001}^{+0.008}$&$10.0_{-2.0}^{+1.8}$&$162_{-31}^{+30}$&4.3&0.10&131.1\\
            &&$/7829/50$&$(6.395-6.404)$&$(7.4-12)$&$(119-206)$&&&\\
NGC 4151(5) &   0.00331900&$701494$&$6.388_{-0.001}^{+0.008}$&$11.1_{-2.5}^{+2.6}$&$44_{-10}^{+10}$&18.4&0.45&63.4\\
            &&$/7830/50.2$&$(6.386-6.397)$&$(7.6-14)$&$(30-59)$&&&\\
NGC 4151(total) &   0.00331900&$\dots$&$6.396_{-0.001}^{+0.001}$&$13.3_{-0.9}^{+0.9}$&$65_{-4}^{+5}$&14.9&0.36&801.5\\
            &&$\dots/389.9$&$(6.395-6.396)$&$(12-14)$&$(59-72)$&&&\\
Mrk 766 &    0.0129290&$700123$&$6.425_{-0.010}^{+0.016}$&$0.8_{-0.5}^{+0.6}$&$37_{-23}^{+27}$&2.3&0.86&7.9\\
            &&$/1597/90.5$&$(6.400-6.450)$&$(0.2-1.6)$&$(9.0-73)$&&&\\
3C 273(1) &     0.158340&$790020$&$6.313_{-0.015}^{+0.013}$&$2.4_{-1.5}^{+1.7}$&$12_{-7}^{+9}$&12.4&811.4&6.7\\
            &&$/459/39.1$&$(6.291-6.334)$&$(0.4-4.8)$&$(2.0-24)$&&&\\
3C 273(2) &     0.158340&$790057$&$6.292_{-0.007}^{+0.007}$&$3.2_{-1.7}^{+1.9}$&$20_{-10}^{+13}$&11.2&734&9.25\\
            &&$/2463/27.1$&$(6.279-6.301)$&$(0.8-6.0)$&$(5.0-38)$&&&\\
3C 273(3) &     0.158340&$790074$&$6.275_{-0.010}^{+0.011}$&$2.5_{-1.5}^{+1.8}$&$22_{-13}^{+15}$&8.4&559&7.2\\
            &&$/3456/25$&$(6.259-6.293)$&$(0.4-5.2)$&$(3.0-45)$&&&\\
3C 273(4) &     0.158340&$790075$&$6.319_{-0.012}^{+0.008}$&$2.2_{-1.7}^{+1.4}$&$20_{-16}^{+12}$&8.3&552&4.3\\
            &&$/3457/25.4$&$\dots$&$(0.0-4.4)$&$(0.0-39)$&&&\\
3C 273(5) &     0.158340&$790076$&$6.412_{-0.029}^{+0.021}$&$1.7_{-1.3}^{+1.6}$&$16_{-12}^{+14}$&8.1&538&4.0\\
            &&$/3573/30.2$&$\dots$&$(0.0-4.1)$&$(0.0-38)$&&&\\
3C 273(6) &     0.158340&$790087$&$6.400^f$&$0_{-0}^{+0.9}$&$0_{-0}^{+7}$&11.9&794&0\\
            &&$/4430/27.6$&$\dots$&$(0.0-1.6)$&$(0.0-10)$&&&\\
3C 273(7) &     0.158340&$790089$&$6.400^f$&$0.0_{-0.0}^{+0.4}$&$0_{-0}^{+3}$&8.0&528&2.0\\
            &&$/5169/30.2$&$\dots$&$(0.0-0.9)$&$(0.0-7.0)$&&&\\
3C 273(total) &     0.158340&$\dots$&$6.319_{-0.013}^{+0.007}$&$1.0_{-0.6}^{+0.6}$&$7_{-4}^{+4}$&9.9&654&8.0\\
            &&$\dots/201.1$&$(6.292-6.328)$&$(0.2-1.8)$&$(1.0-13)$&&&\\
NGC 4593 &   0.00900000&$700279$&$6.399_{-0.008}^{+0.008}$&$2.7_{-0.8}^{+1.0}$&$59_{-18}^{+20}$&4.5&0.81&33.3\\
            &&$/2089/79.9$&$(6.390-6.408)$&$(1.6-4.1)$&$(34-88)$&&&\\
MCG $-$6-30-15(1) &   0.00774900&$700105$&$6.384_{-0.009}^{+0.015}$&$1.4_{-0.6}^{+0.6}$&$28_{-12}^{+12}$&4.7&0.62&15.9\\
            &&$/433/128.2$&$(6.374-6.407)$&$(0.6-2.3)$&$(12-46)$&&&\\
MCG $-$6-30-15(2) &   0.00774900&$700845$&$6.399_{-0.008}^{+0.008}$&$1.2_{-0.4}^{+0.6}$&$26_{-9}^{+12}$&4.5&0.59&17.4\\
            &&$/4759/161.1$&$(6.389-6.409)$&$(0.6-2.1)$&$(17-44)$&&&\\
MCG $-$6-30-15(3) &   0.00774900&$700845$&$6.382_{-0.013}^{+0.009}$&$0.7_{-0.5}^{+0.4}$&$14_{-10}^{+9}$&4.4&0.58&5.5\\
            &&$/4760/172.3$&$(6.359-6.400)$&$(0.0-1.3)$&$(0-27)$&&&\\
MCG $-$6-30-15(4) &   0.00774900&$700845$&$6.400_{-0.018}^{+0.016}$&$0.8_{-0.4}^{+0.6}$&$17_{-8}^{+11}$&4.7&0.63&7.4\\
            &&$/4761/158.8$&$(6.375-6.424)$&$(0.2-1.7)$&$(4.0-33)$&&&\\
MCG $-$6-30-15(5) &   0.00774900&$700845$&$6.342_{-0.021}^{+0.018}$&$1.8_{-1.1}^{+1.3}$&$37_{-22}^{+27}$&4.6&0.61&7.6\\
            &&$/4762/38.2$&$(6.312-6.368)$&$(0.3-3.7)$&$(6.0-77)$&&&\\
MCG $-$6-30-15(total) &   0.00774900&$\dots$&$6.390_{-0.008}^{+0.002}$&$0.8_{-0.2}^{+0.3}$&$18_{-5}^{+5}$&4.5&0.59&29.1\\
            &&$\dots/582$&$(6.375-6.399)$&$(0.5-1.2)$&$(10-25)$&&&\\
IRAS 13349$+$2438(1) &     0.107640&$700902$&$6.400^f$&$0.04_{-0.04}^{+0.1}$&$8_{-8}^{+22}$&0.39&11.5&0.3\\
            &&$/4819/161.9$&$\dots$&$(0.0-0.2)$&$(0.0-46)$&&&\\
IRAS 13349$+$2438(2) &     0.107640&$700902$&$6.428_{-0.008}^{+0.011}$&$0.3_{-0.1}^{+0.3}$&$77_{-31}^{+60}$&0.33&9.7&12.9\\
            &&$/4820/137.5$&$(6.417-6.440)$&$(0.1-0.7)$&$(23-160)$&&&\\
IRAS 13349$+$2438(total) &     0.107640&$\dots$&$6.426_{-0.009}^{+0.013}$&$0.2_{-0.1}^{+0.1}$&$40_{-19}^{+22}$&0.36&10.7&9.7\\
            &&$\dots/294.5$&$(6.411-6.441)$&$(0.0-0.4)$&$(0.0-83)$&&&\\
IC 4329A &    0.0160540&$700367$&$6.399_{-0.005}^{+0.006}$&$3.7_{-1.6}^{+1.6}$&$19_{-8}^{+8}$&17.5&10.1&14.3\\
            &&$/2177/60.1$&$(6.387-6.411)$&$(1.5-6.1)$&$(8.0-31)$&&&\\
Mrk 279 &    0.0304510&$700501$&$6.381_{-0.007}^{+0.008}$&$1.1_{-0.4}^{+0.4}$&$66_{-22}^{+28}$&1.4&2.9&23.4\\
            &&$/3062/116.1$&$(6.372-6.395)$&$(0.5-1.7)$&$(31-107)$&&&\\
NGC 5506 &   0.00618100&$700214$&$6.398_{-0.001}^{+0.008}$&$5.7_{-1.2}^{+1.3}$&$66_{-14}^{+15}$&6.6&0.55&80.6\\
            &&$/1598/90$&$(6.396-6.406)$&$(4.1-7.6)$&$(48-88)$&&&\\
NGC 5548(1) &    0.0171750&$700142$&$6.410_{-0.009}^{+0.016}$&$1.8_{-0.7}^{+0.8}$&$58_{-22}^{+27}$&2.7&1.8&22.1\\
            &&$/837/82.3$&$(6.385-6.434)$&$(0.9-3.0)$&$(29-98)$&&&\\
NGC 5548(2) &    0.0171750&$700485$&$6.394_{-0.007}^{+0.008}$&$1.9_{-0.5}^{+0.5}$&$55_{-14}^{+15}$&3.1&2.0&42.2\\
            &&$/3046/153.9$&$(6.386-6.403)$&$(1.2-2.7)$&$(35-78)$&&&\\
NGC 5548(total) &    0.0171750&$\dots$&$6.402_{-0.010}^{+0.001}$&$1.9_{-0.5}^{+0.4}$&$56_{-14}^{+13}$&2.9&1.9&61.8\\
            &&$\dots/232.7$&$(6.386-6.410)$&$(1.3-2.5)$&$(39-75)$&&&\\
Mrk 290(1) &    0.0295770&$700629$&$6.386_{-0.011}^{+0.012}$&$1.0_{-0.6}^{+0.7}$&$53_{-32}^{+37}$&1.8&3.5&9.0\\
            &&$/3567/55.1$&$(6.367-6.400)$&$(0.2-2.0)$&$(11-106)$&&&\\
Mrk 290(2) &    0.0295770&$700629$&$6.398_{-0.032}^{+0.026}$&$0.5_{-0.4}^{+0.4}$&$36_{-29}^{+28}$&1.3&2.5&3.8\\
            &&$/4399/85.1$&$\dots$&$(0-1.2)$&$(0-85)$&&&\\
Mrk 290(3) &    0.0295770&$700629$&$6.400^f$&$0.4_{-0.3}^{+0.4}$&$20_{-15}^{+22}$&1.8&3.6&1.8\\
            &&$/4441/60.9$&$\dots$&$(0-1.0)$&$(0-52)$&&&\\
Mrk 290(4) &    0.0295770&$700629$&$6.400^f$&$0.3_{-0.3}^{+0.4}$&$15_{-15}^{+21}$&1.8&3.6&0.8\\
            &&$/4442/50.2$&$\dots$&$(0-1.0)$&$(0-51)$&&&\\
Mrk 290(total) &    0.0295770&$\dots$&$6.398_{-0.016}^{+0.009}$&$0.5_{-0.3}^{+0.3}$&$27_{-16}^{+18}$&1.6&3.2&10.8\\
            &&$\dots/247.3$&$(6.374-6.414)$&$(0.2-0.9)$&$(11-50)$&&&\\
PDS 456 &     0.184000&$700742$&$6.400^f$&$0.04_{-0.04}^{+0.06}$&$4_{-4}^{+13}$&0.40&37.3&0.11\\
            &&$/4063/145.2$&$\dots$&$(0-0.2)$&$(0-33)$&&&\\
E1821$+$643 &     0.297000&$700215$&$6.453_{-0.007}^{+0.005}$&$0.7_{-0.3}^{+0.4}$&$26_{-12}^{+13}$&1.4&362.5&13.1\\
            &&$/1599/101.3$&$(6.445-6.463)$&$(0.3-1.3)$&$(11-46)$&&&\\
3C 382(1) &    0.0578700&$700991$&$6.374_{-0.016}^{+0.017}$&$1.0_{-0.8}^{+1.0}$&$16_{-13}^{+15}$&5.5&43.2&3.7\\
            &&$/4910/55$&$\dots$&$(0-2.5)$&$(0-39)$&&&\\
3C 382(2) &    0.0578700&$700991$&$6.408_{-0.010}^{+0.013}$&$1.3_{-0.8}^{+0.9}$&$21_{-13}^{+15}$&4.9&38.5& 7.1\\
            &&$/6151/64.9$&$(6.382-6.429)$&$(0.2-2.6)$&$(3.0-43)$&&&\\
3C 382(total) &    0.0578700&$\dots$&$6.368_{-0.009}^{+0.038}$&$0.9_{-0.6}^{+0.6}$&$14_{-9}^{+9}$&5.2&40.6&6.2\\
            &&$\dots/118$&$(6.351-6.446)$&$(0.1-1.8)$&$(2.0-28)$&&&\\
IRAS 18325$-$5926(1) &    0.0202310&$700587$&$6.400^f$&$0.05_{-0.05}^{+0.35}$&$2_{-2}^{+15}$&2.1&1.9&0.04\\
            &&$/3148/56.9$&$\dots$&$(0-0.7)$&$(0.0-31)$&&&\\
IRAS 18325$-$5926(2) &    0.0202310&$700587$&$6.400^f$&$0.2_{-0.2}^{+0.4}$&$5_{-5}^{+12}$&3.1&2.8&0.2\\
            &&$/3452/51.1$&$\dots$&$(0-1.0)$&$(0.0-29)$&&&\\
IRAS 18325$-$5926(total) &    0.0202310&$\dots$&$6.400^f$&$0.1_{-0.1}^{+0.3}$&$5_{-5}^{+9}$&2.5&2.3&0.2\\
            &&$\dots/106.2$&$\dots$&$(0-0.6)$&$(0-21)$&&&\\
4C $+$74.26(1) &     0.104000&$700679$&$6.258_{-0.014}^{+0.013}$&$1.2_{-0.8}^{+1.0}$&$37_{-25}^{+30}$&2.5&65.1&6.8\\
            &&$/4000/37.7$&$(6.236-6.278)$&$(0.2-2.6)$&$(6.0-79)$&&&\\
4C $+$74.26(2) &     0.104000&$700679$&$6.347_{-0.010}^{+0.011}$&$1.0_{-0.8}^{+1.0}$&$30_{-24}^{+30}$&2.6&67.1&4.7\\
            &&$/5195/31.8$&$(6.322-6.366)$&$(0.0-2.4)$&$(0.0-73)$&&&\\
4C $+$74.26(total) &     0.104000&$\dots$&$6.252_{-0.008}^{+0.011}$&$1.0_{-0.5}^{+0.7}$&$28_{-14}^{+18}$&2.5&66.2&9.5\\
            &&$\dots/66.1$&$(6.242-6.265)$&$(0.3-2.0)$&$(8.0-54)$&&&\\
Mrk 509 &    0.0343970&$700277$&$6.445_{-0.009}^{+0.015}$&$2.2_{-1.0}^{+1.2}$&$34_{-16}^{+17}$&5.8&15.6&13.7\\
            &&$/2087/58.7$&$(6.427-6.462)$&$(0.8-3.9)$&$(12-59)$&&&\\
NGC 7213(1)&   0.00583900&$701410$&$6.395_{-0.008}^{+0.003}$&$2.2_{-0.6}^{+0.6}$&$88_{-23}^{+25}$&2.3&0.18&50.9\\
            &&$/7742/115.3$&$(6.386-6.403)$&$(1.4-3.1)$&$(57-126)$&&&\\
NGC 7213(2)&   0.00583900&$701410$&$6.412_{-0.009}^{+0.016}$&$2.3_{-1.1}^{+1.3}$&$91_{-44}^{+50}$&2.4&0.18&16.5\\
            &&$/8590/35.1$&$(6.395-6.431)$&$(0.9-4.2)$&$(35-164)$&&&\\
NGC 7213(total) &   0.00583900&$\dots$&$6.395_{-0.001}^{+0.008}$&$2.2_{-0.6}^{+0.5}$&$86_{-22}^{+22}$&2.3&0.18&63.3\\
            &&$\dots/150$&$(6.387-6.404)$&$(1.4-3.0)$&$(56-120)$&&&\\
NGC 7314(1) &   0.00474300&$700455$&$6.397_{-0.018}^{+0.015}$&$1.3_{-1.1}^{+1.4}$&$32_{-25}^{+34}$&3.6&0.18&3.8\\
         &&$/3016/28.9$&$\dots$&$(0.0-3.3)$&$(0.0-81)$&&&\\
NGC 7314(2) &   0.00474300&$700455$&$6.422_{-0.009}^{+0.008}$&$1.9_{-0.9}^{+0.9}$&$53_{-22}^{+26}$&3.3&0.16&18\\
            &&$/3719/68.4$&$(6.387-6.437)$&$(0.8-3.2)$&$(23-90)$&&&\\
NGC 7314(total) &   0.00474300&$\dots$&$6.413_{-0.024}^{+0.017}$&$1.5_{-0.8}^{+0.9}$&$41_{-22}^{+24}$&3.4&0.17&15.5\\
            &&$\dots/95.7$&$(6.387-6.437)$&$(0.5-2.8)$&$(19-76)$&&&\\
Ark 564 &    0.0246840&$700168$&$6.400^f$&$0.6_{-0.4}^{+0.4}$&$25_{-16}^{+20}$&2.7&3.7&2.1\\
            &&$/863/49.4$&$\dots$&$(0.0-1.3)$&$(0.0-58)$&&&\\
MR 2251-178 &    0.0639800&$700416$&$6.412_{-0.009}^{+0.008}$&$0.7_{-0.4}^{+0.4}$&$22_{-13}^{+12}$&2.7&25.3&7.7\\
            &&$/2977/148.7$&$(6.396-6.427)$&$(0.1-1.3)$&$(3.0-40)$&&&\\
NGC 7469(1) &    0.0163170&$700395$&$6.388_{-0.007}^{+0.007}$&$2.6_{-0.7}^{+0.8}$&$93_{-24}^{+30}$&2.7&1.6&46.1\\
            &&$/2956/79.9$&$(6.380-6.396)$&$(1.6-3.8)$&$(58-137)$&&&\\
NGC 7469(2) &    0.0163170&$700586$&$6.437_{-0.017}^{+0.008}$&$1.6_{-0.7}^{+0.8}$&$60_{-25}^{+33}$&2.4&1.5&16.7\\
            &&$/3147/69.8$&$(6.412-6.451)$&$(0.6-2.7)$&$(23-104)$&&&\\
NGC 7469(total) &    0.0163170&$\dots$&$6.388_{-0.008}^{+0.002}$&$1.9_{-0.5}^{+0.6}$&$72_{-20}^{+21}$&2.6&1.5&50.9\\
            &&$\dots/147.2$&$(6.379-6.397)$&$(1.2-2.8)$&$(44-104)$&&&\\
\enddata
\tablecomments{
Results from
{\it Chandra } HEG data, fitted with a power law plus Gaussian emission-line model
in the 2--7 keV band, with the line width fixed at 1 eV. All parameters are quoted in the source rest
 frame. 
 Statistical errors are for the 68\% confidence
level, whilst parentheses show the
90\% confidence level ranges of the parameters.
The number of interesting parameters assumed for calculating the
statistical errors was equal to the number of free parameters
in the Gaussian component of the model.
Col.(1): Redshifts obtained from NASA Extragalactic Database  (NED);
Col.(2): Observation sequence number, ID, and exposure time in ks;
Col.(3): Gaussian line centroid energy;
Col.(4): Emission-line intensity in units of $\rm 10^{-5} \ photons \ cm^{-2} \ s^{-1}$;
Col.(5): Emission line equivalent width;
Col.(6): $F$ is the 
estimated 2--10~keV observed flux in units of $10^{-11} \ \rm ergs\ cm^{-2}\ s^{-1}$.
The power-law continuum was extrapolated to 10 keV;
Col.(7): $L$ is the estimated {unabsorbed}
2--10~keV source-frame luminosity
 (using the 2--10 keV estimated flux), in units of $10^{43} \ \rm ergs\ s^{-1}$;
Col.(8): The decrease in the fit statistic, $C$,
when the narrow, two-parameter emission line was added to the 
continuum-only model.  
}
\end{deluxetable}

\begin{deluxetable}{lcccccc}
\tabletypesize{\scriptsize}
\rotate
\tablecaption{\scshape Parameters of the Core Fe K Line Emission ($\sigma$ free) from $Chandra$~(HEG) Data}
\tablewidth{0pt}
\tablehead{
\colhead{Source}  & \colhead{$E^{a}$} & \colhead{$I^{b}$} & \colhead{$EW^{c}$} & \colhead{FWHM$^d$ (Fe~K$\alpha$)}  
& \colhead{FWHM$^d$ (H$\beta)$} &\colhead{Reference$^e$} \\
 \hspace*{10.mm}(1) &  (2) & (3) &  (4) &  (5) & (6) & (7) }
\startdata
Fairall~9  & $6.370^{+0.347}_{-0.161}$ $(6.137-6.906)$ & $5.5^{+13.3}_{-3.6}$  ($1.5-22.9$) & 
$228^{+555}_{-149}$ $(63-954)$ & $18100^{+76840}_{-12390}$  $(5100-121780)$ & 6270$\pm290$ &  N06 \\
Mrk~590 & $6.407^{+0.036}_{-0.033}$  ($6.358-6.461$) & $1.6^{+1.0}_{-0.8}$  ($0.6 - 3.0$) & $171^{+103}
_{-84}$  ($64 - 317$) & $4350^{+6060}_{-2030}$ ($1740 - 15420$) & 2640 & M03 \\
NGC~985  & $6.407^{+0.070}_{-0.076}$  ($6.281 - 6.509$) & $2.2^{+1.6}_{-1.2}$  ($0.6 - 4.4$) 
 & $170^{+127}_{-94}$  ($46 - 344$) & $9550^{+11190}_{-4760}$  ($3810 - 29590$)   & 7500 & M03 \\
ESO~198$-$G24(2)  & $6.385^{+0.013}_{-0.019}$  ($6.353 - 6.404$) & $1.2^{+0.7}_{-0.6}$  ($0.4 - 2.2$)
 & $158^{+105}_{-80}$  ($57 - 304$)     & $<4220$  ($0 - 5840$)  & 6400 & Z05\\
ESO~198$-$G24(total) & $6.382^{+0.025}_{-0.043}$ ($6.306 - 6.426$) & $0.9^{+0.8}_{-0.4}$ ($0.3 - 2.1$)
 & $117^{+110}_{-50}$ ($40-279$)     & $2940^{+8830}_{-1880}$ ($0 - 15500$)  &$\dots$ & $\dots$ \\
3C~120 & $6.410^{+0.016}_{-0.015}$ ($6.389 - 6.439)$ & $3.4^{+1.9}_{-1.5}$ ($1.4 - 6.0$) &
 $66^{+37}_{-30}$ ($27 - 117$)  & $2230^{+2280}_{-1650}$ ($0 - 5950$) & 5370 & W09 \\ 
NGC~2110(1) & $6.389^{+0.098}_{-0.026}$ ($6.341 - 6.521)$& $6.8^{+10.1}_{-3.3}$ ($2.5 - 20.1$)  &
 $116^{+171}_{-56}$ ($43 - 342$) & $4070^{+15260}_{-2470}$ ($0 - 24160$) & 1200$^\star$ & M07 \\
NGC~2110(3) & $6.394^{+0.009}_{-0.007}$ ($6.384 - 6.407)$ & $5.3^{+2.1}_{-1.8}$ ($2.9 - 8.2$) &
 $87^{+35}_{-29}$ ($48 - 135$) & $<2540$ ($0 - 3160$) &$\dots$ &$\dots$ \\
NGC~2110(4)  & $6.395^{+0.010}_{-0.010}$ ($6.381 - 6.409)$ & $5.2^{+1.7}_{-1.5}$ ($3.3 - 7.6$) &
 $127^{+40}_{-37}$ ($80 - 184$)  & $2510^{+2070}_{-1240}$ ($940 - 5600$) & $\dots$&$\dots$\\
NGC~2110(total)  & $6.397^{+0.006}_{-0.006}$ ($6.389 - 6.405)$ & $5.3^{+1.0}_{-1.2}$ ($3.8-6.6$) &
 $103^{+21}_{-23}$ ($75 - 129$) & $2320^{+810}_{-800}$ ($1320-3510$) & $\dots$ &$\dots$\\
PG~0844$+$349(1) & $6.583^{+0.122}_{-0.116}$ $(6.422-6.770)$ & $2.5^{+1.6}_{-1.3}$ $(0.8-4.7)$ 
 & $587^{+384}_{-303}$ $(189-1117)$ & $20320^{+13170}_{-8080}$ $(8900-44490)$ & 2150 & Z05 \\
MCG~$-$5-23-16(1) & 6.384$^{+0.011}_{-0.011}$ $(6.369-6.399)$ & 10.6$^{+3.2}_{-3.1}$ $(6.5-15.0)$
 & 82$^{+25}_{-24}$ $(51-117)$ & 2630$^{+1340}_{-880}$ $(1470-4560)$ & 1450$^\dagger$  & L02 \\
MCG~$-$5-23-16(2)  & 6.408$^{+0.024}_{-0.033}$ $(6.359-6.452)$ & 8.1$^{+5.1}_{-4.1}$ $(2.8-15.1)$ 
 & 65$^{+42}_{-33}$ $(23-122)$ & 3810$^{+4880}_{-1690}$ $(1540-18350)$ & $\dots$& $\dots$\\
MCG~$-$5-23-16(3)  & 6.388$^{+0.019}_{-0.024}$ $(6.352-6.416)$ & 9.6$^{+5.5}_{-4.7}$ $(3.7-18.0)$
& 78$^{+44}_{-38}$ $(30-146)$ & 2660$^{+4150}_{-1580}$ $(610-10420)$ & $\dots$&$\dots$\\
MCG~$-$5-23-16(total) & 6.388$^{+0.009}_{-0.009}$ $(6.377-6.400)$ & 9.0$^{+2.1}_{-2.2}$ $(6.1-12.1)$
 & 71$^{+17}_{-17}$ $(48-96)$ &  2560$^{+1130}_{-900}$ $(1390-4180)$ & $\dots$&$\dots$ \\
NGC~3516(1)  & 6.392$^{+0.005}_{-0.006}$ $(6.385-6.399)$ & 3.9$^{+1.2}_{-1.3}$ $(2.3-5.6)$
 & 110$^{+34}_{-36}$ $(65-158)$ & $<1670$ $(0-3160)$ & 3353$\pm310$ & P04 \\
NGC~3516(2)  & 6.408$^{+0.010}_{-0.011}$ $(6.393-6.422)$ & 4.4$^{+2.0}_{-1.7}$ $(2.2-7.2)$
 & 186$^{+85}_{-71}$ $(93-306)$ & 1740$^{+1420}_{-1210}$ $(0-4020)$ &$\dots$& $\dots$ \\
NGC~3516(3)  &  6.402$^{+0.017}_{-0.014}$ $(6.382-6.425)$ & 4.5$^{+1.6}_{-1.4}$ $(2.7-6.7)$
 & 157$^{+56}_{-49}$ $(94-234)$ & 4290$^{+2180}_{-1470}$ $(2450-8050)$ &$\dots$ &$\dots$ \\
NGC~3516(4)  &  6.409$^{+0.023}_{-0.025}$ $(6.374-6.442)$ & 4.2$^{+3.1}_{-2.0}$ $(1.6-8.3)$
 & 67$^{+51}_{-31}$ $(26-134)$ & 3220$^{+3020}_{-1630}$ $(960-8190)$ & $\dots$& $\dots$\\
NGC~3516(5)  &  6.354$^{+0.057}_{-0.079}$ $(6.241-6.431)$ & 5.1$^{+4.1}_{-2.6}$ $(1.7-10.6)$
 & 101$^{+81}_{-52}$ $(34-209)$ & 8480$^{+7050}_{-4700}$ $(3110-18840)$ &$\dots$ &$\dots$\\
NGC~3516(6)  &  6.407$^{+0.033}_{-0.034}$ $(6.364-6.451)$ & 7.6$^{+3.7}_{-3.4}$ $(3.3-12.6)$
 & 108$^{+52}_{-49}$ $(47-178)$ & 6030$^{+3550}_{-2430}$ $(2970-11240)$ &$\dots$ &$\dots$\\
NGC~3516(7)  &  6.414$^{+0.017}_{-0.017}$ $(6.389-6.437)$ & 4.2$^{+2.4}_{-2.0}$ $(1.7-7.5)$
 & 55$^{+32}_{-26}$ $(23-99)$ & 2290$^{+2150}_{-1310}$ $(460-5900)$ &  $\dots$ &$\dots$\\
NGC~3516(total)  &  6.404$^{+0.007}_{-0.007}$ $(6.395-6.413)$ & 4.4$^{+0.8}_{-0.7}$ $(3.4-5.5)$
 & 91$^{+17}_{-14}$ $(71-114)$ & 3180$^{+880}_{-670}$ $(2310-4390)$ & $\dots$ &$\dots$\\
NGC~3783(1)  & 6.396$^{+0.014}_{-0.013}$ $(6.377-6.415)$ & 5.4$^{+2.7}_{-2.6}$ $(2.3-9.1)$ &
69$^{+36}_{-33}$ $(30-118)$ & $<4670$ $(0-5780)$ & 3570$\pm$190 & N06 \\
NGC 3783(2)  & 6.401$^{+0.006}_{-0.0063}$ $(6.392-6.410)$ & 5.1$^{+1.5}_{-1.1}$ $(3.6-7.1)$ &
75$^{+21}_{-16}$ $(53-104)$ & 1930$^{+1080}_{-900}$ $(750-3490)$ & $\dots$&$\dots$ \\
NGC 3783(3)  & 6.391$^{+0.008}_{-0.008}$ $(6.380-6.401)$ & 6.2$^{+1.4}_{-1.5}$ $(4.3-8.2)$ &
90$^{+22}_{-21}$ $(63-120)$ & 2700$^{+1180}_{-1050}$ $(1410-4430)$ &$\dots$ &$\dots$  \\
NGC 3783(4)  & 6.395$^{+0.005}_{-0.006}$ $(6.388-6.402)$ & 6.0$^{+1.3}_{-1.4}$ $(4.2-7.8)$ &
88$^{+19}_{-21}$ $(62-114)$ & 1860$^{+880}_{-1140}$ $(0-3140)$ &  $\dots$& $\dots$\\
NGC 3783(5)  & 6.395$^{+0.004}_{-0.005}$ $(6.388-6.401)$ & 6.3$^{+1.5}_{-1.1} (4.8-8.3)$ &
69$^{+16}_{-12}$ $(53-91)$ & 1280$^{+720}_{-630}$ $(0-2260)$ & $\dots$& $\dots$\\
NGC 3783(6)  & 6.399$^{+0.006}_{-0.006}$ $(6.391-6.408)$ & 4.7$^{+1.2}_{-1.2}$ $(3.2-6.4)$ &
57$^{+14}_{-15}$ $(39-77)$ & 1520$^{+890}_{-940}$ $(0-2750)$ &$\dots$ &$\dots$\\
NGC 3783(total)  & 6.396$^{+0.003}_{-0.002}$ $(6.393-6.399)$ & 5.6$^{+0.5}_{-0.6}$ $(4.8-6.3)$ &
74$^{+7}_{-8}$ $ (63-83)$ & 1750$^{+360}_{-360}$ $(1270-2240)$ & $\dots$& $\dots$\\
NGC 4051  & 6.417$^{+0.039}_{-0.036}$ $ (5.750-6.474)$ & 3.5$^{+1.4}_{-1.4}$ $ (1.6-5.5)$ &
195$^{+79}_{-78}$ $ (89-307)$ & 6430$^{+11800}_{-2860}$ $ (2840-479470)$ & 1200 & W09\\
NGC 4151(1)  & 6.396$^{+0.006}_{-0.006}$ $(6.386-6.404)$ &  21.7$^{+3.3}_{-4.1}$ $(16.5-26.3)$ 
 & 190$^{+29}_{-36}$ $(146-231)$ & 2150$^{+1220}_{-680}$ $(1250-3840)$ & 6350 & W09\\
NGC 4151(2)  & 6.391$^{+0.004}_{-0.004}$ $(6.386-6.397)$ &  18.2$^{+2.7}_{-2.5}$ $(14.9-21.8)$
 & 78$^{+11}_{-11}$ $(64-93)$ & 2170$^{+610}_{-540}$ $(1460-3000)$ & $\dots$& $\dots$\\
NGC 4151(3)  & 6.396$^{+0.006}_{-0.005}$ $(6.389-6.404)$ &  20.3$^{+3.6}_{-3.5}$ $(15.7-25.3)$
 & 83$^{+15}_{-14}$ $(64-103)$ & 2670$^{+790}_{-680}$ $(1760-3770)$ & $\dots$& $\dots$\\
NGC 4151(4)  & 6.400$^{+0.006}_{-0.005}$ $(6.393-6.408)$ &  11.5$^{+2.9}_{-2.6}$ $(8.1-15.5)$
 & 188$^{+47}_{-43}$ $(132-253)$ & 1710$^{+860}_{-740}$ $(690-2940)$ & $\dots$&$\dots$ \\
NGC 4151(5)  & 6.393$^{+0.016}_{-0.008}$ $(6.382-6.416)$ &  14.3$^{+8.3}_{-4.2}$ $(8.9-25.4)$
 & 57$^{+33}_{-17}$ $(36-101)$ & 2020$^{+3600}_{-870}$ $(420-6750)$ & $\dots$&$\dots$ \\
NGC 4151(total)  & 6.394$^{+0.003}_{-0.002}$ $(6.391-6.398)$ &  17.5$^{+1.6}_{-1.5}$ $(15.5-19.7)$
 & 87$^{+8}_{-8}$ $(77-98)$ & 2250$^{+400}_{-360}$ $(1770-2790)$ & $\dots$ &$\dots$\\
3C 273(1)     & 6.336$^{+0.074}_{-0.053}$ $(6.259-6.491)$ & 5.9$^{+5.0}_{-4.0}$ $(0.6-13)$ &
  35$^{+29}_{-24}$ $(4-76)$ & 5900$^{+8640}_{-5830}$ $(0-80630)$ & 3520 & Z05 \\
NGC 4593 & 6.406$^{+0.011}_{-0.042}$ $(6.351-6.421)$ & 3.8$^{+3.4}_{-1.4}$ $(2.0-8.5)$ & 
82$^{+74}_{-30}$ $(43-185)$ & 2230$^{+8180}_{-1100}$ $(670-15320)$ & 3650 & W09\\
MCG~$-$6-30-15(1)  & 6.399$^{+0.043}_{-0.045}$ $(6.335-6.461)$ & 3.4$^{+1.5}_{-1.8}$ $(1.0-6.1)$
 & 70$^{+31}_{-37}$ $(21-125)$  & 7440$^{+5710}_{-4630}$ $(2480-16270)$ & 1700$\pm$170 & N06 \\ 
MCG~$-$6-30-15(2)  & 6.395$^{+0.061}_{-0.044}$ $(6.298-6.485)$ & 1.9$^{+1.4}_{-0.9}$ $(0.7-3.9)$
 & 40$^{+30}_{-19}$ $(15-83)$  & 2810$^{+15450}_{-1750}$ $(0-23770)$ &$\dots$ &$\dots$\\
MCG~$-$6-30-15(3)  & 6.424$^{+0.094}_{-0.110}$ $(6.237-6.557)$ & 2.8$^{+2.1}_{-1.5}$ $(0.8-5.7)$
 & 62$^{+45}_{-34}$ $(17-124)$  & 13590$^{+12820}_{-6050}$ $(5760-37720)$ &$\dots$ &$\dots$\\
MCG~$-$6-30-15(4)  & 6.402$^{+0.154}_{-0.023}$ $(6.364-6.602)$& 1.0$^{+1.0}_{-0.7}$ $(0.1-2.3)$
& 20$^{+20}_{-14}$ $(2-45)$ & $<14800$ $(0-28590)$ & $\dots$ &$\dots$\\
MCG~$-$6-30-15(5) & 6.345$^{+0.027}_{-0.024}$ $(6.292-6.393)$ & 2.2$^{+2.3}_{-1.4}$ $(0.3-5.3)$
 & 46$^{+48}_{-29}$ $(6-111)$  &$<5850$ $(0-15200)$  & $\dots$ &$\dots$ \\
MCG~$-$6-30-15(total)  & 6.427$^{+0.044}_{-0.044}$ $(6.366-6.486)$ & 2.7$^{+1.1}_{-0.9}$ $(1.5-4.1)$
 & 58$^{+23}_{-20}$ $(32-88)$  & 11880$^{+4650}_{-4030}$ $(6480-18750)$ & $\dots$ & $\dots$  \\
IRAS~13349$+$2438(2)  & 6.396$^{+0.046}_{-0.057}$ 
& 0.7$^{+0.5}_{-0.4}$ $ (0.2-1.5)$ & 170$^{+139}_{-93}$ $(52-386)$ & 5870$^{+11370}_{-2550}$ $(2660-85550)$ 
& $\dots$&$\dots$ \\
IRAS~13349$+$2438(total) & 6.405$^{+0.052}_{-0.113}$ $(6.106-6.841)$ & 0.4$^{+0.5}_{-0.2}$ $(0.07-2.3)$ 
& 87$^{+121}_{-41}$ $(16-532)$& 5150$^{+60200}_{-2810}$ $(1770-93230)$ & $\dots$ & $\dots$  \\
IC~4329A  & 6.305$^{+0.139}_{-0.096}$ $(6.172-6.542)$ & 15.8$^{+10.6}_{-7.5}$ $(5.9-31.5)$
 & 81$^{+54}_{-38}$ $(30-162)$ & 18830$^{+18590}_{-9620}$ $(5820-48080)$ & 5620$\pm200$ & N06\\
Mrk 279 &  6.414$^{+0.054}_{-0.028}$ $(6.312-6.560)$ & 2.0$^{+1.2}_{-0.9}$ $(0.9-6.9)$ &
132$^{+80}_{-59}$ $(60-458)$ & 5080$^{+8390}_{-1940}$ $(2670-48780)$ & 5150 & W09 \\
NGC 5506  & 6.400$^{+0.007}_{-0.006}$ $(6.391-6.409)$ & 7.1$^{+1.6}_{-2.1}$ $(4.4-9.3)$
 & 84$^{+18}_{-25}$ $(52-109)$ & 1650$^{+880}_{-870}$ $(470-2940)$ & 1850 & Z05\\
NGC 5548(1)  & 6.398$^{+0.022}_{-0.021}$ $(6.367-6.427)$ & 3.7$^{+1.5}_{-1.4}$ $(1.9-5.8)$
 & 124$^{+51}_{-46}$ $(64-195)$ & 4410$^{+2590}_{-1580}$ $(2390-8500)$ & 5830$\pm230$ & N06 \\
NGC 5548(2)  & 6.402$^{+0.009}_{-0.009}$ $(6.389-6.415)$ & 2.4$^{+0.9}_{-0.7}$ $(1.4-3.6)$
 & 71$^{+27}_{-21}$ $(42-107)$ & 1960$^{+1040}_{-900}$ $(800-3540)$ & $\dots$ &$\dots$ \\
NGC 5548(total) & 6.403$^{+0.009}_{-0.009}$ $(6.391-6.415)$ & 2.7$^{+0.8}_{-0.7}$ $(1.8-3.7)$
 & 84$^{+25}_{-22}$ $(56-115)$ & 2540$^{+1140}_{-820}$ $(1490-4240)$ & $\dots$ &$\dots$\\
Mrk 290(total)  & 6.404$^{+0.037}_{-0.038}$ $(6.342-6.458)$ & 1.0$^{+0.6}_{-0.5}$ $(0.2-1.9)$
 & 60$^{+33}_{-31}$ $(12-110)$ & 5290$^{+5120}_{-2420}$ $(2140-20200)$ & 4740 & W09 \\
E1821$+$643  & 6.447$^{+0.051}_{-0.054}$ $(6.355-6.517)$ & 3.2$^{+1.6}_{-1.3}$ $(1.6-5.4)$
& 153$^{+75}_{-63}$ $(76-257)$   & 10920$^{+7710}_{-3910}$ $(5640-25950)$ & 6620$\pm720$ & N06\\
3C 382(2)   & 6.424$^{+0.064}_{-0.090}$ $(6.254-6.531)$ & 3.6$^{+3.2}_{-2.3}$ $(0.7-8.1)$ 
& 66$^{+55}_{-43}$ $(13-144)$ & 8100$^{+12580}_{-4490}$ $(3150-33320)$ & 8340 & W09 \\
3C 382(total)   & 6.418$^{+0.084}_{-0.097}$ $(6.161-6.538)$ & 3.4$^{+2.3}_{-2.5}$ $(0.4-7.1)$
& 57$^{+39}_{-42}$ $(7-119)$ & 10730$^{+15810}_{-8320}$ $(2060-55310)$ & $\dots$ &$\dots$ \\
4C $+$74.26(total)  & 6.260$^{+0.038}_{-0.081}$ $(6.125-6.392)$ & 1.2$^{+2.0}_{-0.8}$ $(0.1-4.1)$
 & 36$^{+60}_{-24}$ $(3-124)$  & $<10980$ $(0-23620)$ & 9420 & W09 \\
Mrk 509    & 6.428$^{+0.020}_{-0.021}$ $(6.396-6.455)$ & 3.6$^{+2.0}_{-1.7}$ $(1.4-6.4)$
 & 57$^{+30}_{-27}$ $(22-100)$   & 2910$^{+2590}_{-1250}$ $(1280-7900)$ & 3430$\pm240$ & N06 \\
NGC 7213(1)  & 6.392$^{+0.013}_{-0.011}$ $(6.377-6.410)$ & 2.9$^{+1.1}_{-1.0}$ $(1.6-4.4)$
 & 117$^{+46}_{-40}$ $(65-179)$   & 2290$^{+1950}_{-1390}$ $(390-5000)$ & 3200 & Z05 \\
NGC 7213(2)  & 6.410$^{+0.018}_{-0.018}$ $(6.384-6.436)$ & 3.2$^{+2.0}_{-1.6}$ $(1.2-6.0)$
 & 126$^{+78}_{-63}$ $(47-236)$   & 2400$^{+2310}_{-1800}$ $(0-6770)$ &  $\dots$&$\dots$ \\
NGC 7213(total)  & 6.397$^{+0.011}_{-0.010}$ $(6.384-6.412)$ & 3.0$^{+1.0}_{-0.9}$ $(1.8-4.3)$
 & 120$^{+42}_{-35}$ $(73-174)$   & 2590$^{+1470}_{-1170}$ $(1050-4620)$ &$\dots$ &$\dots$ \\
NGC 7469(1)  & 6.385$^{+0.010}_{-0.012}$ $(6.364-6.399)$ & 3.2$^{+1.1}_{-1.2}$ $(1.7-5.1)$
 & 116$^{+42}_{-42}$ $(63-187)$   & 1800$^{+2640}_{-1360}$ $(0-6040)$ & 2820 & W09\\
NGC 7469(2)  & 6.395$^{+0.036}_{-0.033}$ $(6.347-6.452)$ & 3.9$^{+2.0}_{-1.6}$ $(1.8-6.6)$
 & 156$^{+79}_{-64}$ $(72-262)$   & 6780$^{+4810}_{-3170}$ $(3200-14100)$ &  $\dots$&$\dots$\\
NGC 7469(total)  & 6.388$^{+0.018}_{-0.017}$ $(6.365-6.413)$ & 3.7$^{+1.0}_{-1.1}$ $(2.2-5.2)$
 & 142$^{+38}_{-42}$ $(84-199)$   & 4890$^{+2770}_{-1700}$ $(2740-8740)$ &$\dots$ &$\dots$ \\
\enddata
\tablecomments{ \small
Results from
{\it Chandra} HEG data, fitted with a power law plus Gaussian emission-line model 
in the 2--7 keV band.
Statistical errors are for the 68\% confidence
level, whilst parentheses show the
90\% confidence level ranges of the parameters.
$^{\star}$  Broad polarized H$\beta$ line.
$^{\dagger}$  Infra-red broad Br$\alpha$ line.
$^{a}$  Gaussian line center energy in keV.   
$^{b}$  Emission-line intensity in units of $\rm 10^{-5} \ photons \ cm^{-2} \ s^{-1}$.
$^{c}$  Emission line equivalent width in units of eV.
$^{d}$  Full width half maximum, rounded to $10 \ \rm km \ s^{-1}$.
$^{e}$  References for H$\beta$ FWHM:  L02$-$Lutz et al. (2002); M03$-$Marziani et al. (2003);
 M07$-$Moran et al. (2007); N06$-$Nandra (2006); P04$-$Peterson et al. (2004); W09$-$Wang et al. (2009); 
Z05$-$Zhou et al. (2005).
}
\end{deluxetable}

\begin{deluxetable}{lcccc}
\tablecaption{\scshape 
Mean \fekalfa Line Spectral Parameters}
\tablewidth{0pt}
\tablehead{
\colhead{Parameter}  & \colhead{By Observation} & \colhead{\# Spectra$^{a}$} &
 \colhead{By Source} & \colhead{\# Sources$^{b}$}} 
\startdata

Centroid Energy (keV)    &   $6.396 \pm 0.0004$ $^{c}$  & 68 &  $6.397 \pm 0.0005$ $^{c}$ & 32 \\
($\sigma_{\rm Fe~K}$ fixed) & & & & \\

Centroid Energy  (keV)  &    $6.388 \pm 0.001$  & 68 &  $6.398 \pm0.002$ &  32 \\
($\sigma_{\rm Fe~K}$ free)$^{d}$ & & & & \\

EW   (eV)              &    $42 \pm 2$  &       70 &     $44 \pm 2$ & 33 \\
($\sigma_{\rm Fe~K}$ fixed) & & & & \\

EW (eV) &   $53 \pm 3$  &           70  &    $70 \pm 4$ & 33 \\
($\sigma_{\rm Fe~K}$ free)$^{d}$ & & & & \\

FWHM ($ \rm km \ s^{-1}$) & $2060 \pm 230$     &   53   &   $2200 \pm 220$ & 27 \\
\enddata
\tablecomments{ \small
Weighted mean quantities from spectral fitting to
individual spectra (``by observation''), and to spectra
representative of each source (``by source''). See text, \S\ref{properties}
for details. $^{a}$ Number of spectra (one per observation) contributing
to the mean quantities. $^{b}$ Number of unique sources 
contributing to the mean quantities.
$^{c}$ These statistical errors are smaller than the systematic
errors (see \S\ref{centroidenergy} for discussion).
$^{d}$ The intrinsic width of the \fekalfa in these cases was
free in the spectral fitting for 51 spectra in the individual
observation fits (see \tablesigfreep) and for 27 sources in
the fits to source-representative spectra (see \S\ref{properties}
for details). 
}
\end{deluxetable}

\begin{deluxetable}{lcc}
\tablecaption{\scshape 
X-ray Baldwin Effect: \fekalfa Line EW versus Luminosity Fits}
\tablewidth{0pt}
\tablehead{
\colhead{Parameter}  & \colhead{By Observation} &
 \colhead{By Source} }
\startdata

$\log{EW} = k_{L} + [m_{L} \log{(L_{x})}]$ & & \\

$\chi^{2}$ (d.o.f.)    &         58.5(68)        &                 24.7(31) \\

intercept, $k_{L}$ &   $1.80^{+0.02}_{-0.02}$    & $1.76^{+0.02}_{-0.02}$ \\

slope, $m_{L}$ (68\% confidence errors)  &      $-0.22^{+0.03}_{-0.03}$  
	&      $-0.13^{+0.04}_{-0.04}$ \\

slope,  $m_{L}$ (99\% confidence errors)  &    $-0.22^{+0.10}_{-0.07}$  
	&       $-0.13^{+0.11}_{-0.11}$ \\

$\Delta \chi^{2}$   for $m_{L}=0$ &     39.3      &          10.0 \\

Significance for
$m_{L} \neq 0$       &   $6.27\sigma$       &           $3.24\sigma$ \\

& & \\
$\log{EW} = k_{R} + [m_{R} \log{(L_{x}/L_{\rm EDD})}]$ & & \\

$\chi^{2}$ (d.o.f.)     &           61.4 (68)    &                25.2(31) \\

intercept, $k_{R}$ (68\% confidence errors)      & $1.31^{+0.09}_{-0.09}$ 
	&             $1.50^{+0.09}_{-0.09}$ \\

slope, $m_{R}$ (68\% confidence errors) &  $-0.20^{+0.03}_{-0.03}$ &  
$-0.11^{+0.04}_{-0.04}$ \\

slope, $m_{R}$ (99\% confidence errors) &   
	$-0.20^{+0.07}_{-0.11}$   &  $-0.11^{+0.09}_{-0.09}$ \\

$\Delta \chi^{2}$  for $m_{R}=0$      &      36.6            &               9.5 \\

Significance for $m_{R} \neq 0$  &  $6.05\sigma$  &       $3.08\sigma$ \\
\enddata
\tablecomments{ \small
Results of fitting the relations between the derived \fekalfa line
EWs and the 2--10~keV X-ray luminosity ($L_{x}$), and between
the EWs and the Eddington ratio ($L_{x}/L_{\rm EDD}$). Coefficients 
and their error bounds are shown for linear fits to $\log{EW}$ versus
$\log{(L_{x})}$ ($k_{L}$, $m_{L}$), and to $\log{EW}$ versus
$\log{(L_{x}/L_{\rm EDD})}$ for spectral fitting results to the
individual spectra (``by observation''), and to the source-representative
spectra. See 
\S\ref{baldwineffect} for details.
The number of degrees of freedom for each fit (d.o.f.) is shown in
parentheses after each best-fitting $\chi^{2}$ value.
}
\end{deluxetable}

\end{document}